\documentclass[prb,twocolumn,showpacs,floatfix,superscriptaddress]{revtex4}
\usepackage{graphicx,epsfig}
\begin{document}
\title{Electron correlation effects in electron-hole recombination in 
organic light-emitting diodes}
\author{Kunj Tandon$^\dagger$}
\footnote{$^\dagger$Present address: GE John F. Welch Technology Center,
Sy 152, Export Promotion Industrial Park, Ph 2, Hoodi Village, 
White Field Road, Bangalore 560 066, India}
\affiliation{Solid State and Structural Chemistry Unit, 
Indian Institute of Science, Bangalore 560 012, India}
\author{S. Ramasesha}
\affiliation{Solid State and Structural Chemistry Unit, 
Indian Institute of Science, Bangalore 560 012, India}
\author{S. Mazumdar}
\affiliation{Department of Physics, and The Optical Sciences Center,
University of Arizona, Tucson, AZ 08721}
\date{\today}
\begin{abstract}
We develop a general theory of electron--hole recombination in organic 
light emitting diodes that leads to formation of emissive singlet excitons 
and nonemissive triplet excitons. 
We briefly review other existing theories and show how our approach is
substantively different from these theories.
Using an exact time-dependent approach to the 
interchain/intermolecular charge--transfer 
within a long--range interacting model
we find that, (i) 
the relative yield of the singlet exciton in polymers is considerably 
larger than the 25\% predicted from statistical considerations,
(ii) the singlet exciton yield increases with chain length in oligomers,
and, (iii) in small molecules containing nitrogen heteroatoms, the
relative yield of the singlet exciton is considerably smaller and may be
even close to 25\%. The above results are independent of whether or not
the bond-charge repulsion, 
$X_{\perp}$,
is included in the interchain part of the
Hamiltonian for the two-chain system. The larger (smaller) 
yield of the singlet (triplet)
exciton in carbon-based long-chain polymers is a consequence of both its
ionic (covalent) nature and smaller (larger) binding energy. 
In nitrogen containing monomers,
wavefunctions are closer to the noninteracting limit, and this decreases
(increases) the relative yield of the singlet (triplet) exciton.
Our results are in qualitative agreement with electroluminescence 
experiments involving both molecular and polymeric light emitters. The
time-dependent approach developed here for describing intermolecular
charge-transfer processes is completely general and may be applied to many
other such processes.
\end{abstract}
\maketitle
\section{Introduction}
\label{intro_section}

Charge recombination and photoinduced charge-transfer lie at the heart of
current attempts to construct viable optoelectronic devices using organic
semiconducting materials consisting of $\pi$-conjugated polymers or 
molecules. Charge recombination
is the fundamental process of interest in organic light emitting diodes 
(OLEDS). Electroluminescence (EL) in OLEDS results from,
(a) the injection
of electrons and holes into thin films containing the emissive material,
(b) migration of these charges, which can involve both coherent motion
on a single chain and interchain or intermolecular charge-transfer between
neutral and charged species, (c) 
recombination
of electrons and holes on the same polymer chain or molecule
\cite{Tang87a,Burroughes90a,Gustafsson92a}. If the 
recombination leads to the singlet optical exciton, light emission can occur.
If, on the other hand, the final product of the recombination is a triplet
exciton, only nonradiative relaxation can occur in the absence of strong
spin-orbit coupling. EL in OLEDS is of strong current interest, both because
of applications in display devices \cite{Garten97a,friend99a} 
and the potential for obtaining organic
solid state lasers \cite{Dodabalapur}. 
The fundamental process that occurs in
photoinduced charge-transfer is the exact reverse of that in EL:
optical excitation to the singlet
exciton in a donor molecule is followed by charge separation and migration
of charge to a neighboring acceptor molecule. The latter process is of 
interest in photovoltaic applications \cite{Sari99a}.

The fundamental electronic process of charge recombination or
separation is therefore of strong current interest.
Especially in the context of EL in OLEDS, charge recombination
has received both
experimental and theoretical attention (see below). The overall
quantum efficiency of the EL depends on, (i) the fraction of the total 
number of injected carriers that end up as excitons
on the same polymeric chain or molecule, (ii) the fraction of these 
excitons that are spin singlets, since only singlet
excitons are emissive, and (iii) the fraction of singlet excitons that 
actually undergo radiative decay.
In the present paper we focus on (ii), which determines the maximum possible
EL efficiency.

Formally, the charge recombination process can be written as,
\begin{equation}
\label{CT}
P^+ + P^- \to G + S/T
\end{equation}
where $P^{\pm}$ are charged polaronic states of the emissive molecule, $G$
is the ground state of the neutral molecule, and $S$ and $T$ are singlet and
triplet excited states of the neutral molecule. Eq.~\ref{CT} indicates that
both singlet and triplet excitons are likely products of the charge 
recombination process. We shall denote the fraction of singlet excitons
generated in OLEDS by the above recombination process as $\eta$.

Early discussions of $\eta$ were based on statistical arguments alone.
Since electrons and holes are injected independently from the two electrodes,
and since two spin-1/2 particles can give three independent spin 1 states
(with $M_S$ = --1, 0 and +1) but only one spin 0 state ($M_S$ = 0), it 
follows that $\eta$ is 0.25. 
Note, however, that this argument is strictly valid only
for noninteracting electrons, such that single-configuration molecular 
orbital descriptions of all eigenstates are valid.
In such a case, the highest occupied and
lowest unoccupied molecular orbitals (HOMO and LUMO) are identical for the
singlet and triplet excited states. Charge recombination (Eq.~\ref{CT})
then involves merely the
migration of an electron from the doubly (singly) occupied HOMO (LUMO) of
$P^-$ to the singly occupied (unoccupied) HOMO (LUMO) of the $P^+$, for 
both singlet and triplet channels. The singlet channel and all three 
triplet channels of the charge recombination process are equally likely 
within the MO scheme. If electrons are interacting, however, this 
simple single-configuration description breaks down, as all the states
included in Eq.~\ref{CT} are now superpositions of multiple configurations.
There is no longer any fundamental reason for the singlet and triplet 
channels to be equally likely processes, and hence there is no reason 
for $\eta$ to be 0.25.

Experimentally, $\eta$ has been found to range from $\sim$ 0.25 -- 0.66
\cite{Baldo99a,Cao99a,Ho00a,Wohlgenannt01a} in different materials. In 
OLEDS with the molecular species Aluminum tris (8-hydroxyquinoline)
(Alq$_3$) as the emissive material Baldo et. al.  have determined 
$\eta \sim$ 0.22 $\pm$ 0.03, in agreement with that expected from 
statistical arguments \cite{Baldo99a}. On the other hand, considerably
larger $\eta \sim$ 0.45 has been found in derivatives of 
poly(para-phenylenevinylene) (PPV) by Cao et al. and Ho et. al.
\cite{Cao99a,Ho00a}. Wohlgenannt et. al., using spin-dependent recombination
spectroscopy, have determined the formation cross-sections of singlet and
triplet excitons, $\sigma_S$ and $\sigma_T$, respectively, 
for a large number of polymeric materials (including nonemissive 
polymers in which the lowest two-photon state 2A$_g$ occurs below the 
optical 1B$_u$ exciton), and found that $\sigma_S/\sigma_T$ is strongly 
material dependent and in all cases considerably larger than 1 (thereby 
implying that $\eta$ is material dependent and much larger than 
0.25) \cite{Wohlgenannt01a}. More recently, Wilson et al.
\cite{Wilson01a} and Wohlgenannt et. al. \cite{Wohlgenannt02a} have shown
that $\eta$ can depend strongly on the effective conjugation length, with 
values ranging from $\sim$ 0.25 for small monomers to considerably larger
than 0.25 for long chain oligomers.      

Theoretically, $\eta$ has been investigated by a number of groups
\cite{Wohlgenannt01a,Hong01a,Kobrak00a,Shuai00a,Ye02a} including 
ourselves. There is general agreement that $\eta$ can be substantially 
greater than 0.25 in $\pi$-conjugated polymers and that this is an electron
correlation effect. There exist, however, substantial differences between
the assumptions and formalisms that go into these theories. The goal of the
present work is to develop a formalism that gives a clear physical picture
of the electron-hole recombination and explains why $\eta$ substantially 
larger than 0.25 is to be 
expected in organic polymeric systems. Ideally, since photoinduced
charge-transfer is the exact reverse process of electron-hole recombination,
it should also be possible to extend our approach to 
photoinduced charge-transfer in the future.
Brief presentation of our work has been made earlier \cite{Wohlgenannt01a},
where, however, the emphasis was more on the experimental technique used by
our experimental collaborators. Here we present the full theoretical details 
of our earlier work, provide a critique of the earlier theories and also 
report on the new and interesting
results of our investigation of external electric field effects on $\eta$, 
albeit for artificially large fields, and also on the role of nitrogen
heteratoms in electron--hole recombination.
Specifically, our theoretical approach involves a time-dependent formalism, 
within which the initial state composed of two oppositely charged polarons is 
allowed to propagate in time under the influence of the complete Hamiltonian 
that includes both on-chain and interchain interactions. For the sake of 
completeness, we also discuss other existing theoretical approaches 
\cite{Hong01a,Kobrak00a,Shuai00a,Ye02a}, and their applicability to real 
systems. In particular, there exists a superficial similarity between the 
approach used in references \onlinecite{Shuai00a}, \onlinecite{Ye02a} and 
ours. For a physical understanding of the electron-hole recombination process 
it is essential (see below) that the difference between our approach and that 
used by the authors of reference \onlinecite{Shuai00a}, \onlinecite {Ye02a} 
is precisely understood.

The plan of the paper is as follows. In section \ref{model}, we present
our theoretical models for intrachain and interchain interactions, and also
discuss the model systems that are studied. In section
\ref{theories} we present a brief critique of the existing theories.
A more extended 
discussion of the approach used by Shuai et. al. \cite{Shuai00a,Ye02a}
is given in Appendix 1. In section 
\ref{prop} we present the method of propagation of the initial state, while
in section \ref{results} we present our numerical results. In this section
we also discuss an alternate approach to the time propagation
for the simplest case of two ethylenes that confirms the validity of the 
more general approach used in section \ref{prop}, and 
that also gives a physical picture of the recombination
process. While $\eta >$ 0.25 is found in our calculations with interacting
electrons, the absolute yields of 
both singlet and triplet excitons are found to be
extremely small with standard electron correlation parameters. We therefore
investigate the effects of the external electric field on these yields
within a highly simplified model. It is found that for sufficiently large
fields the yields with interacting electrons are as large as those with
noninteracting electrons in the field-free case, and that in the relatively
small field region $\eta$ continues to be greater than 0.25. In the very high
field regime it is found that $\eta$ can even be smaller than 0.25. 
While the bare electric fields required to see the reversal of the 
singlet-triplet ratio are rather large and therefore only of academic 
interest, if internal field effects are taken into account, it is possible
to envisage situations where the effective electric field is large 
enough to bring about such a reversal in the singlet-triplet ratio.
Following the discussion of electric field effects, we discuss how the
chain length dependence of $\eta$, as observed experimentally 
\cite{Wilson01a,Wohlgenannt02a}, can be understood within our theory.
We then consider the role of heteroatoms, especially in the context
of molecular emitters. We show that in small systems with heteroatoms
$\eta$ can approach the statistical limit, thus explaining qualitatively 
the monomer results of Wilson et. al. \cite{Wilson01a}, and the
results of Baldo et. al. for Alq$_3$ \cite{Baldo99a}.
The emphasis in all our calculations is on understanding the qualitative
aspects of charge recombination and not on detailed quantitative aspects.
Finally in section \ref{conclusion} we 
discuss the conclusions and scope of future work.

\section{Theoretical model}  
\label{model}

The goal of the present work is to provide benchmark results for the
charge recombination reaction which are
valid for the strong Coulomb interactions that characterize $\pi$-conjugated
systems. Accurate treatments of electron-electron interactions are not
possible for long chain systems, and in this initial study we have therefore
chosen pairs of short polyene chains, with 2 -- 6 carbon atoms in each chain 
as our model systems. Since polyene eigenstates possess mirror-plane and
inversion symmetries, we shall henceforth refer to the ground state
$G$ (see Eq.~\ref{CT}) as 1$^1A_g$, and $S$ and $T$ as 1$^1B_u$ and 1$^3B_u$, 
respectively. The model system containing two hexatrienes (12 carbon atoms 
overall) is the largest system that can be treated exactly at present
within correlated 
electron models.  

Our approach suffers from two apparent disadvantages. First, 
polyenes and polyacetylenes are weakly emissive because the 2$^1A_g$ state in 
these occur below the optical 1$^1B_u$ state. This presents no problem as far 
as the analysis of the EL in emissive materials is concerned, as the 
spectroscopic technique of Wohlgenannt et. al. \cite{Wohlgenannt01a} find 
a strong deviation of $\sigma_S/\sigma_T$ from 1 even in systems with energy 
ordering similar to that in polyenes
\cite{Wohlgenannt01a} (see results for PTV in this paper, for instance), 
and as we show in the following, this is a direct consequence of the 
large energy difference between the singlet 1$^1B_u$ exciton
and the triplet 1$^3B_u$ exciton, as well as the fundamental difference in
their electronic structures. Both, in turn, are consequences of
strong electron-electron interactions, which 
also characterize systems like PPV and poly-paraphenylene (PPP),
as evidenced from the large difference in energies between the 
singlet and triplet excitons in these systems, determined experimentally
\cite{Monkman99a,Monkman01a,Osterbacka,Bassler}, as well as theoretically
\cite{Beljonne,Chandross}. 
A second apparent disadvantage of our procedure is related to the limitation
of our calculations  
to short systems. This prevents direct evaluation of the chain length
dependence of $\eta$. 
We believe that this problem can be 
circumvented once the mechanism of the physical process that leads to
the difference between singlet and triplet generation is precisely understood,
and for this purpose it is essential that the electron correlation effects
are investigated thoroughly using exactly solvable models. As we show later, 
our approach gives a precise though qualitative explanation of the chain 
length dependence.

Our model system consists of two polyene chains of equal lengths
that lie directly on top of each other, separated by 4 \AA. 
We consider the charge recombination process of Eq.~\ref{CT}, and there
are two possible initial states: (i) a specific chain (say chain 1)
is positively charged,
with the other (chain 2) having negative charge, a configuration that 
hereafter we denote as $P^+_1P^-_2$, where the subscripts 
1 and 2 are chain indices, or (ii) 
the superposition $P^+_1P^-_2 \pm P^+_2P^-_1$, in the same notation. In our 
calculations we have chosen the first as the proper initial state, since
experimentally in the OLEDS the symmetry between the chains
is broken by the external electric
field (we emphasize that the consequence of choosing the symmetric or 
antisymmetric superposition can be easily predicted from our all our 
numerical calculations that follow).
Even with initial state (i),
the final state can consist of both $(1^1A_g)_1(1^1B_u)_2$ and
$(1^1A_g)_2(1^1B_u)_1$ in the singlet channel. The same is true in the
triplet channel, i.e., either of the two chains can be in 
the ground (excited) state.
Hereafter we will write the initial states as $|i_S\rangle$ and
$|i_T\rangle$, where the subscripts $S$ and $T$ correspond to spin states
$S$ = 0 and 1. We consider only the $M_S$ = 0 triplet state.
The initial states are simply the product states with 
appropriate spin combinations,

\begin{eqnarray}
\label{initial_states}
|i_S\rangle = 2^{-1/2}(|P^+_{1,\uparrow}\rangle |P^-_{2,\downarrow}\rangle -
|P^+_{1,\downarrow}\rangle |P^-_{2,\uparrow}\rangle) \\
 \nonumber \\
|i_T\rangle = 2^{-1/2}(|P^+_{1,\uparrow}\rangle |P^-_{2,\downarrow}\rangle +
|P^+_{1,\downarrow}\rangle |P^-_{2,\uparrow}\rangle)
\end{eqnarray}

There exist of course two other initial triplet states with $M_S = \pm$ 1.
The overall Hamiltonian for our composite two-chain system consists
of an intrachain terms $H_{intra}$ and  
interchain interactions $H_{inter}$. Additional interactions must be 
explicitly included to discuss external influences like the electric 
field etc.  $H_{intra}$ describing individual chains is the 
Pariser-Parr-Pople (PPP) Hamiltonian \cite{Pariser53a,Pople53a}
for $\pi$-electron systems, written as,
\begin{eqnarray}
H_{intra} = -\sum_{<ij>,\sigma}t_{ij} (a^{\dagger}_{i,\sigma}
a^{}_{j,\sigma} + H.C.) + \sum_i \epsilon_i n_i + \nonumber \\
\sum_i U_i n_{i,\uparrow}n_{i,\downarrow} 
+  \sum _{i>j} V_{ij} (n_i -z_i) (n_j -z_j)  
\label{PPP}
\end{eqnarray}
where $a^{\dagger}_{i,\sigma}$ creates a $\pi$-electron of spin $\sigma$ on
carbon atom $i$, $n_{i,\sigma} = a^{\dagger}_{i,\sigma}a^{}_{i,\sigma}$ is 
the number of electrons on atom $i$ with spin $\sigma$ and 
$n_i = \sum_{\sigma}n_{i,\sigma}$ is the total number of electrons on atom 
$i$, $\epsilon_i$ is the site energy and $z_i$ are the local chemical 
potentials. The hopping matrix element 
$t_{ij}$ in the above are restricted to nearest neighbors and 
in principle can contain electron-phonon interactions, although 
a rigid bond approximation is used here. $U_i$ and $V_{ij}$ are the
on-site and intrachain intersite Coulomb interactions. 

We use standard parameterizations for $H_{intra}$. The hopping integrals for
single and double bonds are taken to be 2.232 eV and 2.568 eV, respectively
and all the site energies of carbon atoms in a polymer with all equivalent 
sites are set to zero.
We choose the Hubbard interaction parameter $U_C$ for carbon to be 11.26 eV, 
and for the
$V_{ij}$ we choose the Ohno parameterization \cite{Ohno64a},
\begin{eqnarray}
V_{ij} = 14.397\left[\left({{28.794}\over {U_i+U_j}}\right)^2~+~r^2_{ij}
\right]^{-{{1} \over {2}}}
\label{Ohno}
\end{eqnarray}
where the distance $r_{ij}$ is in \AA, ~$V_{ij}$ is in eV and the local 
chemical potential $z_C$ for $sp^2$ carbon is one. It should be noted then
when hetero atoms like nitrogen are present, the on-site correlation energy,
the site energy and the local chemical potential could be different from 
those for carbon. For $H_{inter}$, we choose the following form,
\begin{eqnarray}
\label{inter}
H_{inter} = -t_{\perp}\sum_{i,\sigma}(a^{\dagger}_{i \sigma}
a^{\prime}_{i,\sigma} + H.C.) + \nonumber \\
+ X_{\perp}\sum_{i,\sigma}(n_i + n^{\prime}_i)(a^{\dagger}_{i \sigma}
a^{\prime}_{i,\sigma} + H.C.) + \nonumber \\
\sum_{i,j} V_{i,j}(n_i -z_i)(n^{\prime}_j - z_{j^\prime})
\end{eqnarray}
In the above, primed and unprimed operators correspond to sites on different 
chains. Note that the interchain hopping $t_{\perp}$ is restricted to 
corresponding sites on the two chains, which are nearest interchain neighbors.
The interchain Coulomb interaction $V_{i,j}$, however, includes interaction 
between any site on one chain with any other site on the other chain.
In addition to the usual one-electron hopping that occurs within the zero 
differential overlap approximation \cite{Pariser53a,Pople53a} we have also 
included a many-electron site charge-bond charge repulsion $X_{\perp}$ 
(operating between nearest interchain neighbors only) that consists of 
multicenter Coulomb integrals. This term should also occur within $H_{intra}$, 
but is usually ignored there because of its small magnitude, relative to all 
other terms \cite{Pariser53a,Pople53a,Campbell90a}. In contrast, the 
$t_{\perp}$ in $H_{inter}$ is expected to be much smaller, and $X_{\perp}$ 
cannot be ignored in interchain processes, especially at large interchain 
separations \cite{Rice96a}. We have done calculations for both $X_{\perp}$
= 0 and $X_{\perp} \neq$ 0.

\section{Brief Critique of Existing theories}
\label{theories}

To put our work in the proper context we present a discussion of the existing 
theories of charge recombination \cite{Hong01a,Kobrak00a,Shuai00a,Ye02a} in 
this section. The natures of
$H_{intra}$ within all these models are similar in the sense that they
all incorporate intrachain Coulomb interactions, without which of course
there cannot be any difference between singlet and triplet generation. 
Following this, there is a fundamental difference between the models of
references \onlinecite{Hong01a}, \onlinecite{Kobrak00a} on the one hand, 
and those of references \onlinecite{Shuai00a}, \onlinecite{Ye02a} and ours 
on the other. Within the theory of references \onlinecite{Hong01a}, 
\onlinecite{Kobrak00a}, there is no difference in singlet or triplet
generation in the first stage of the charge-recombination process,
which involves {\it interchain} charge-transfer.
Within these models, interchain charge-transfer yields high
energy singlet and triplet excited states of long chains
that occur in the continuum, and the lowest singlet and triplet excitons
result from relaxations of these high energy states. Differences
in the relative yields of the lowest singlet and triplet excitons are
consequences of differences in the {\it intrachain relaxation processes}
in the singlet and triplet channels, that occur in the second stage of the
overall process. In contrast, within our theory \cite{Wohlgenannt01a} and
the theory of references \onlinecite{Shuai00a} and \onlinecite{Ye02a}, the 
lowest singlet and triplet excitons are generated directly from two 
oppositely charged polarons, and their different yields are
consequences of the different cross-sections of the {\it interchain}
charge-transfer reactions in the singlet and triplet channels.

Within the model of Hong and Meng \cite{Hong01a}, the continuum singlet state
decays to the lowest singlet exciton, while the continuum triplet state
decays to a high energy triplet state $T_2$ consisting of a loosely bound 
triplet exciton, which then relaxes nonradiatively to the lowest tightly 
bound triplet exciton $T_1$.  The energy gap between $T_2$ and $T_1$ is large, 
and according to Hong and Meng, this 
nonradiative relaxation has to be a multiphonon cascade process. The large
energy gap and the multiphonon nature of the relaxation creates a 
``bottleneck''
in the $T_2 \to T_1$ 
nonradiative transition, and spin-orbit coupling leads to intersystem 
crossing from 
$T_2$ to the singlet exciton, thereby increasing the relative yield of
singlets \cite{Hong01a}. 
We believe that the key problem with this approach is that the
model is in disagreement with what is known about the spectrum of triplet
states from triplet absorptions in $\pi$-conjugated polymers 
\cite{Osterbacka} and theoretical solutions to the PPP model 
\cite{Chandross}. 
Experimentally, in PPV, for instance, the lowest triplet occurs at about 1.55 
eV \cite{Osterbacka}, while in MEH-PPV this state occurs at $\sim$ 1.3 eV
\cite{Monkman99a}. The triplet absorption energy in these systems is
about 1.4 eV. Theoretically,
the final state in triplet absorption occurs slightly below the continuum band
\cite{Chandross}, and this is therefore the $T_2$ state (also referred to as
the $m^3A_g$ \cite{Osterbacka}). The energy region
between $T_2$ and $T_1$ (m$^3A_g$ and 1$^3B_u$)
in the triplet subspace is not at all
sparse, as assumed by Hong and Meng,
but rather, within the correlated PPP Hamiltonian
$H_{intra}$ in Eq. \ref{PPP}, this energy region contains numerous other
triplet states \cite{Ramasesha84a,Tavan87a}.
Thus any nonradiative
relaxation from $T_2$ to $T_1$ in the realistic systems should involve a 
number of intermediate
triplet states with small energy gaps between them, 
and therefore the phonon bottleneck simply
will not occur. An additional problem with the model of Hong and Meng
is that even in the
singlet channel, generation of the lowest exciton from a continuum singlet
state cannot be direct but can
occur only through the $m^1A_g$ loosely bound singlet exciton 
\cite{Chandross}. In principle, this can lead to a bottleneck even in the
singlet channel. To summarize, we believe that the model of Hong and Meng
is in disagreement with the known singlet and triplet energy spectra within
the PPP model.

Within the model of Kobrak and Bittner \cite{Kobrak00a}
also polaron pairs are formed on the
single chain first. These authors take into account the electron-phonon
interactions explicitly, and the two-particle states on a single chain
are allowed to evolve by interacting with a one-dimensional classical
vibrational lattice. Different cross-sections for singlet and
triplet excitons are found within the authors' model, and the difference
originates from the difference in the mixing between the polaron and exciton
states with different spin. The theory includes only the Coulomb interactions
between the polaron charges and not the Coulomb interactions between all
the electrons that appear in the PPP Hamiltonian. The theory also assumes 
large quantum efficiency for the generation of the high energy states with
the two polaron charges on the same chain, starting from a state with the
charges on different chains. A recent calculation by Ye et.
al. \cite{Ye02a} indicates very weak cross-sections for the generation of
high energy $^1B_u$ and $^3B_u$
states starting from the initial state containing the
charges on different chains (see Fig.~8 in reference \onlinecite{Ye02a}). 
This is supported also by our exact calculations (see below). 
However, the calculations by Ye et al. \cite{Ye02a} as well as ours are for 
relatively short chains, and further work is needed to test the validity of 
the model of Kobrak and Bittner. 
As we show in section \ref{results}, $\eta >$ 0.25 is predicted 
from considerations of the initial stage of interchain charge-transfer alone. 
Whether additional contributions can come from differences in the intrachain 
relaxation processes needs to be studied further.

We now come to the work by Shuai et. al. \cite{Shuai00a,Ye02a}, who, like us,
have determined $\eta >$ 0.25 in oligomers of PPV from considerations of
interchain charge-transfer. Precisely because of the apparent similarity of 
our approaches, it is essential that we discuss the approach of Shuai et. al.
in detail, since our ultimate goal is to arrive at a physical explanation of
the greater yield of the singlet exciton than what is predicted from 
statistical considerations, and as we show later, the physical mechanisms
within references \onlinecite{Shuai00a}, \onlinecite{Ye02a} and within our 
work are quite different. The quantity that is calculated in references 
\onlinecite{Shuai00a}, \onlinecite{Ye02a} is $\sigma_S/\sigma_T$, viz., the 
ratio of the formation cross-sections of the 1$^1B_u$ singlet and 1$^3B_u$ 
triplet exciton. For fast spin-lattice interaction, 
the expression for $\eta$ in terms of $\sigma_S$ and 
$\sigma_T$ can be written as \cite{Wohlgenannt01a,Wohlgenannt02a},
\begin{equation}
\eta = \sigma_S/(\sigma_S + 3\sigma_T)
\label{eta}
\end{equation}
and thus, for $\sigma_S/\sigma_T >$ 1, $\eta >$ 0.25.

Shuai et. al. consider the same $H_{intra}$ as us, and $H_{inter}$ that is 
similar (see below).  The authors then use the Fermi ``Golden Rule'' approach 
to calculate $\sigma_S$ and $\sigma_T$. 
According to the authors, the cross-section ratio is given by,
\begin{equation}
\label{fermi}
\sigma_S / \sigma_T = 
|\langle i_S|H_{inter}|f_S \rangle|^2/|\langle i_T|H_{inter}|f_T \rangle|^2
\end{equation}
where $|i_S\rangle$  and $|i_T\rangle$ are the singlet and triplet 
initial states (see Eqs. 2 and 3), and
$|f_S\rangle$ and $|f_T\rangle$ are the corresponding final states, 
respectively. Since the interchain Coulomb interaction is
diagonal in the space of the states considered in Eq.~\ref{fermi}, the
authors ignore $V_{i,j}$ in Eq.~\ref{inter} but retain the other terms.
Shuai et. al. find that for $X_{\perp}$ = 0 in Eq.~\ref{inter}, 
when the interchain charge-transfer is
due to the hopping $t_{\perp}$ only, the right hand side of Eq.~\ref{fermi}
is $\sim$ 1, a result we agree with (see Appendix 1). The authors then 
claim that for nonzero positive $X_{\perp}$, and for positive 
$t_{\perp}$ (note negative sign in front of the one-electron term in 
Eq.~\ref{inter}), the right hand side of Eq.~\ref{fermi}
can be substantially larger than 1. The authors calculated the matrix 
elements in Eq.~\ref{fermi}
for pairs of PPV oligomers in parallel configuration using approximate
methods (singles configuration interaction \cite{Shuai00a} and 
coupled-cluster method \cite{Ye02a}), 
and have found the right hand side of Eq.~\ref{fermi} 
to show divergent behavior over a broad 
range of $X_{\perp}/t_{\perp}$ (see Fig. 1 in reference \onlinecite{Shuai00a} 
and Figs. 3, 4, 6 and 7 in reference \onlinecite{Ye02a}).
Based on these calculations the authors conclude that a moderate to large
$X_{\perp}$ is essential for the experimentally observed large $\sigma_S
/\sigma_T$ \cite{Cao99a,Ho00a,Wohlgenannt01a,Wilson01a,Wohlgenannt02a}.

This result is surprising, in view of the fact that the site charge-bond charge
repulsion is {\it spin-independent}, exactly as the one-electron interchain
hopping in Eq.~\ref{inter}.
Since this question is intimately linked with the mechanism of charge
recombination that we are after we have re-examined this issue by performing 
exact calculations for pairs of polyene chains with lengths N = 2, 4 and 6. 
The conclusions from these exact calculations are described below.

As discussed above, even with $P^+_1P^-_2$ as the initial state (with, of
course, appropriate spin functions) the final state contains two terms,
with one of the two chains in the ground state and the other in the
excited state. Instead of working with different superpositions of the
final states we consider $\sigma_S$ to be proportional to 
$|\langle i_S|H_{inter}|(1^1A_g)_1(1^1B_u)_2\rangle|^2$ +
$|\langle i_S|H_{inter}|(1^1A_g)_2(1^1B_u)_1\rangle|^2$.
Similarly, $\sigma_T$ is taken to be proportional to
$|\langle i_T|H_{inter}|(1^1A_g)_1(1^3B_u)_2\rangle|^2$ +
$|\langle i_T|H_{inter}|(1^1A_g)_2(1^3B_u)_1\rangle|^2$.
As shown explicitly in the Appendix, the magnitudes of the matrix 
elements of the initial singlet [triplet] $P^+_1P^-_2$ with
$(1^1A_g)_1(1^1B_u)_2$ [$(1^1A_g)_1(1^3B_u)_2$] and $(1^1A_g)_2(1^1B_u)_1$
[$(1^1A_g)_2(1^3B_u)_1$] are different for $X_{\perp} \neq$ 0, and hence
the final states cannot be 1:1 superpositions of these configurations.
Note that by taking the sums of the squares we exhaust all possibilities 
automatically.
For the conclusions of 
references \onlinecite{Shuai00a}, \onlinecite{Ye02a} to be valid the 
calculated $\sigma_S/\sigma_T$ within Eq.~\ref{fermi} should now show 
strong dependence on $X_{\perp}/t_{\perp}$ (as mentioned above divergent 
$\sigma_S/\sigma_T$ is implied in references \onlinecite{Shuai00a,Ye02a}).
Our exact results for the three different chain lengths are shown in Fig. 1
below, where we see that only for $X_{\perp}/t_{\perp}$ 
very close to 0.5 is $\sigma_S/\sigma_T$, as calculated within 
Eq.~\ref{fermi}, is substantially different from 1. 
At all other $X_{\perp}/t_{\perp}$ the RHS of Eq.~\ref{fermi}
is very close to 1.
Furthermore, except for $X_{\perp}/t_{\perp}$ = 0.5 the chain length 
dependence of $\sigma_S/\sigma_T$ is weak. If we now recall that all chain 
length dependent quantities (for example, optical and other energy gaps in 
polyenes \cite{Ramasesha84a}) exhibit strongest length dependence
at the shortest 
lengths, the conclusion that emerges is that except for the unique point
$X_{\perp}/t_{\perp}$ = 0.5, $\sigma_S/\sigma_T$ remains $\sim$ 1 within 
the Golden Rule approach even in the long chain limit.

In order to understand this difference from the results of Shuai et. al.
\cite{Shuai00a,Ye02a} in further detail we present analytic results for the
case of two ethylenes (N = 2) in Appendix 1. These results are important in 
so far as they begin to give a physical picture for the charge recombination
reaction, even as they indicate that the site charge-bond charge repulsion
is not the origin of large $\eta$. The analytic calculations also make 
the origin of the uniqueness of the point $X_{\perp}/t_{\perp}$ = 0.5 
absolutely clear. Indeed it is seen that precisely at this point both  
$\sigma_S$ and $\sigma_T$, as defined in Eq. ~\ref{fermi}, approach zero. 
More importantly,
the chain length-independence, as suggested in Fig. 1
can be understood very clearly from the analytic calculations. 
Finally, it can also be seen from these
calculations that had we taken the initial state to be the superposition
$P^+_1P^-_2 \pm P^+_2P^-_1$, instead of only one of these, the 
$\sigma_S/\sigma_T$, as calculated from Eq. ~\ref{fermi}
would be exactly 1 for all $X_{\perp}/t_{\perp}$. 

Our 
basic conclusion then is that the Fermi Golden Rule approach is not valid 
for calculations of $\sigma_S/\sigma_T$ or $\eta$. This is to be expected 
also from a different consideration, viz., the Fermi Golden Rule approach 
is valid for calculations of states that lie within a narrow band, whereas 
in the present case the energy difference between the initial and final 
states, and that between the singlet and triplet excitons are both much 
larger than $t_{\perp}$ and $X_{\perp}$. The origin of the difference between 
our exact calculations of matrix elements and the approximate calculations 
of Shuai et. al. is harder to ascertain. 
One possibility is that the polaron wavefunctions are open shell, and 
approximating these within mean field or limited CI could lead to wrong
conclusions.

In the following sections we therefore go beyond the Fermi Golden Rule 
approach to understand the origin of large $\eta$.

\section{Time evolution of the polaron pair state}
\label{prop}

A straightforward numerical solution of ${H_{intra} + H_{inter}}$ will 
merely give the electronic structure of the composite two-chain system. 
Such a calculation does not contain any information about the relative 
yields of specific final states starting from the initial two-polaron 
states. Our approach therefore consists of propagating the initial state 
in time under the influence of the complete Hamiltonian, and monitoring 
the time-evolved state to obtain information about the final products. 

In principle, given a Hamiltonian, propagation of any initial 
state is easily achieved by solving the
time-dependent Schrodinger equation. One could use the interaction picture
to separate the nontrivial evolution of the initial state from the trivial
component which occurs as a result of the evolution of the product of the
eigenstates of the Hamiltonian of the subsystems \cite{Messiah72a}. 
In the context of the many-body
PPP Hamiltonian such an approach is difficult to implement numerically. This
is because the total number of eigenstates for the two-chain system is very
large: the number of such states for two chains of six carbon atoms each is
853,776 in the $M_s$ = 0 subspace. Obtaining all the eigenstates of the
two-component system and expressing the matrix elements of $H_{inter}$ in the
basis of these eigenstates is therefore very intensive computationally. It is
simpler to calculate the time evolution in the Schrodinger representation,
determine the time-evolved states, and project them on to the desired
final eigenstates (for instance, $|1^1A_g\rangle_1|1^1B_u\rangle_2$). This 
is the approach we take.

We first obtain the eigenstates $|P^+_1\rangle$, $|P^-_2\rangle$ as well as 
the product states exactly in the valence bond (VB) basis \cite{Ramasesha84a}
(in which the total spin $S$ is a good quantum number) in
order to avoid spin contamination. Following the time-evolution, however,
we need to calculate overlaps of the time-evolved states with various final
states (see below), which is cumbersome within the 
nonorthogonal VB basis. After calculating the exact spin singlet and
triplet initial states, we therefore expand these in  
an orthonormal basis that has only well defined total $M_S$  value.

Henceforth we refer to the initial states  
$|i_S\rangle$ and $|i_T\rangle$ collectively as
$\Psi(0)$ and the 
time-evolved states as $\Psi(t)$. In principle, the time evolution can be
done by operating on $\Psi(0)$ with the time evolution operator,
\begin{equation}
U(0,t) = \exp(-iHt)
\label{time}
\end{equation}
where $H$ is the total Hamiltonian $H_{intra} + H_{inter}$. This approach 
would, however, require obtaining a matrix representation of the exponential
time evolution operator, which in turn requires the determination of the
prohibitively large number of eigenstates of the composite two-chain system.
We can avoid this problem by using small discrete time intervals and expanding 
the exponential operator in a Taylor series, and stopping at the linear term.
Such an approach, however, has the undesirable effect of spoiling unitarity, 
and for long time evolutions would lead to loss of normalization of the 
evolved state.
The way around this dilemma has been proposed and used by others
\cite{Crank47a,Varga62a} in different contexts and involves using the 
following truncated time-evolution scheme,
\begin{eqnarray}
(1 + iH{{\Delta t} \over {2}}) \Psi(t+\Delta t) =
(1 - iH{{\Delta t} \over {2}}) \Psi(t)
\label{evolve}
\end{eqnarray}
In the above equation, on the left hand side, we evolve the state at time 
$(t+\Delta t)$ backwards by $\Delta t/2$ while on the right hand side,
we evolve the state at time $t$ forward by $\Delta t/2$. By forcing these
two to be equal, we ensure unitarity in the time evolution of the state.
It can be seen easily that this time evolution which is accurate to 
${{\Delta t^2} \over {2}}$ is unitary. For a given many-body Hamiltonian 
and initial state, the right hand side of Eq.~\ref{evolve} is a vector 
in the Hilbert space of the two-chain Hamiltonian.  The left hand side 
corresponds to the action of a matrix on an as yet unknown vector, that 
is obtained by solving the above set of linear algebraic equations. 
Further details of the numerical procedure can be found in Appendix 2.

After each evolution step, the evolved state is projected onto the space of
neutral product eigenstates of the two-chain system. The relative yield
$I_{mn}(t)$ for a given product state $|m,n\rangle$ = $|m\rangle_1|n\rangle_2$
is then obtained from,
\begin{equation}
I_{mn}(t) = |\langle\Psi(t)|m,n \rangle|^{2}
\label{overlap}
\end{equation}
In our case the states $|m,n\rangle$ can be any of the 
final states of interest,
viz., $|(1^1A_g)_1(1^1B_u)_2\rangle$, $|(1^1A_g)_1(1^3B_u)_2\rangle$, etc.
It is for efficient calculations of the overlaps (while at the same time
maintaining spin purity) in Eq.~\ref{overlap} that we expand our exact 
eigenstates of the neutral system in the VB basis to the total 
$M_S$ basis. We emphasize that $I_{mn}(t)$ is a measure of the yield of the 
state $|m,n\rangle$ at time $t$ and is not a cross-section.

\section{Numerical Results}
\label{results}

In this section we report the results of our calculations of recombination
dynamics for for pairs of ethylenes, butadienes and hexatrienes, both within
the noninteracting H\"uckel model ($U_i = V_{ij} = X_{\perp} = 0$)
and the interacting PPP model. Following
this, we show the results of our investigation of electric field effects
on the same systems,
discuss the chain length dependence of $\eta$, and finally present the
numerical results for a model system containing nitrogen heteroatoms.
The calculations for the noninteracting case provides a check of our numerical
procedure,
and the comparison between the noninteracting and the interacting
model allows us to determine the effect of electron-electron interactions.

\subsection{Dynamics in the H\"uckel Model}

While there is no difference in energy between singlets and triplets in the
H\"uckel Model, it is nevertheless possible to have spin singlet and triplet
initial states $|i_S\rangle$ and $|i_T\rangle$, as well as
singlet and triplet final states. In Fig. 2
we show the yield 
for the electron-hole recombination in the singlet channel, for pairs of
ethylenes, butadienes and hexatrienes. The yields for the triplet channels
are not shown separately in this case, -- we have ascertained that these 
are identical to those
in the singlet channel in this case, as expected. 
These calculations are for $t_{\perp}$ = 0.1 eV within Eq.~\ref{inter}.
We note that the yields $I_{mn}(t)$ oscillate with time.
This is to be expected within our purely electronic Hamiltonian, within 
which an electron or hole jumps back and forth between the two molecular 
species. These
oscillations are the analogs of the Rabi oscillations \cite{Rabi37a,Allen87a}
that occur
upon the stimulation of a system with light, where absorption of light can
occur only with nonzero damping. Within our purely electronic Hamiltonian,
complete transition to the final states can only occur in the
presence of damping (for
example, radiative and nonradiative relaxations of the final states), that
has not been explicitly included in our Hamiltonian. The frequency of 
oscillation is higher for larger intermolecular transfer integral 
$t_{\perp}$, as expected. The frequency of the oscillation also depends 
upon the size of the molecule and is lower for larger molecules (see below
for an explanation of this). 
The equalities in the yields of the singlet and triplet excited states
found numerically conforms to the simple free spin
statistics which predicts that in the $M_S=0$ state formed from electron-
hole recombination, the probability of singlet and triplet formation are
equal. Since the $M_S=\pm 1$ cases always yield triplets, the spin
statistics corresponding to 25\% singlets and 75\% triplets is
vindicated in this case.

Although the H\"uckel calculations do not yield any new information, 
it is useful
to pursue them further in order to arrive at a physical mechanism of the
charge recombination process. To this end we have developed an alternate
procedure for calculating the above dynamics for the smallest model system,
viz., a pair of ethylenes. This alternate approach
consists of expanding the initial state $\Psi(0)$ as a superposition 
of the eigenstates $\psi_i$ of the composite two-chain system
with eigenvalues $E_i$, 
\begin{eqnarray}
 |\Psi(0)\rangle =\sum_i c_i|\psi_i(0)\rangle
\end{eqnarray}
The evolution of the state $\Psi(0)$ is now simply given by
\begin{eqnarray}
 |\Psi(t)\rangle =\sum_i c_i |\psi_i(0)\rangle \exp (-iE_it/\hbar)
\end{eqnarray}
The yield $I_{mn}(t)$ in a given channel with final state $|m,n\rangle$
is then obtained from, 
\begin{eqnarray}
I_{mn}(t) = |\langle m,n|\Psi(t)\rangle|^2 
~~~~~~~~~~~~~~~~~~~~~~~~~~~~~~~~~~~~~~~\nonumber
 \\
= \sum_i |c_i \langle m,n|\psi_i(0)\rangle \exp(-iE_it/\hbar)|^2
~~~~~~~~~~~~~ \nonumber \\
= \sum_i |c_i\langle m,n|\psi_i(0)\rangle|^2 + 
~~~~~~~~~~~~~~~~~~~~~~~~~~~\nonumber \\
~~~~\sum_i \sum_{j > i} 2~{\rm Re}~\left\{c_i c_j\langle m,n|\psi_i(0)\rangle 
\langle \psi_j(0)|m,n\rangle\right\}  \nonumber \\
\times~{\rm cos} ((E_i-E_j)t/\hbar)~~~~~~~~~~~~~~~~~~~~~~~
\label{yld_t}
\end{eqnarray}
The quantities $\langle m,n|\psi_i(0) \rangle$ are readily obtained from the
eigenstates of the neutral one-chain subsystems and the composite eigenstates 
of the two-chain
system. In Tables I and II we list the nonzero values of the coefficients
$c_i$ and the $\langle m,n|\psi_i(0) \rangle$ 
values for the case of two
\begin{table}
\caption[]{Significant $c_i = < \Psi(0)|\psi_i >$ and the
$< m,n|\psi_i >$ values and their product
in the H\"uckel model for a pair ethylenes in singlet channel.
The index $i$ corresponds to the index of `significant' eigenstates of the 
total system and $E_i$ the corresponding energy eigenvalue.}
\label{tbl_huc_cff_s}
\begin{center}
\begin{tabular}{|c|c|c|c|c|} \hline
 ~i~  & $E_i$ (eV) & $c_i$ & $< m,n|\psi_i >$  &
$ <m,n|\psi_i > c_i $ \\  \hline
2 & -4.3360 & 0.3691 & 0.1362 & 0.0503 \\
3 & -4.3360 & -0.3373 & 0.1138 & -0.0384 \\
4 & -4.1360 & -0.0171 & 0.0120 & -0.0002 \\
5 & -4.1360 & -0.5000 & 0.0058 & -0.0029 \\
6 & -4.1360 & 0.4989 & 0.0120 & 0.0057 \\
7 & -4.1360 & -0.0285 & 0.0059 & -0.0002 \\
8 & -3.9360 & -0.3558 & 0.1266 & -0.0450 \\
9 & -3.9360 & 0.3513 & 0.1234 & 0.0433 \\ \hline
\end{tabular}
\end{center}
\end{table}
\begin{table}
\caption[]{ Significant $c_i = < \Psi(0)|\psi_i > $ and
$ < m,n|\psi_i > $,
for the triplet channel, for a pair of ethylenes. }
\label{tbl_huc_cff_t}
\begin{center}
\begin{tabular}{|c|c|c|c|c|} \hline
~i~  & $E_i$ (eV) & $ < \Psi(0)|\psi_i > $ &
$< m,n|\psi_i > $  &
$< m,n|\Psi_i > c_i$   \\  \hline
2 & -4.3360 & 0.3373 & 0.1138 & 0.0384 \\
3 & -4.3360 & 0.3691 & 0.1362 & 0.0503 \\
4 & -4.1360 & -0.0179 & 0.0037 & -0.0001 \\
5 & -4.1360 & 0.4985 & 0.0180 & 0.0090 \\
6 & -4.1360 & 0.5005 & 0.0047 & 0.0024 \\
7 & -4.1360 & 0.0251 & 0.0170 & 0.0004 \\
8 & -3.9360 & 0.3513 & 0.1234 & 0.0433 \\
9 & -3.9360 & 0.3558 & 0.1266 & 0.0450 \\ \hline
\end{tabular}
\end{center}
\end{table}
ethylenes. It is seen that sets of degenerate states of the composite system
together contribute equally to the singlet and triplet channels, although
individual members of the set
may contribute unequally. We have determined that the 
time evolution obtained from this approach is exactly the same
as that obtained from the general method described in the previous section.

The contribution arising from the right hand side of Eq.~\ref{yld_t} has
been separated
into time-independent and time-dependent parts. The latter comes
about whenever the two eigenstates in question are nondegenerate. 
Furthermore, at $t$ = 0 the contribution from the time independent
part exactly cancels the contribution from the time dependent part. 
When the sign of the time dependent part becomes positive the two 
contributions add up to give the maximum yield of 0.25 in both the 
singlet and the triplet channels observed in the discrete calculations.
The periodicity of the oscillation corresponds to the energy
difference between the two pairs of the degenerate states. This analysis
could in principle be extended to the case of the larger systems but 
would be quite tedious in view of the larger Hilbert space dimensions.
Note that the decrease of the oscillation frequency of $I_{mn}(t)$ with 
increasing chain
length (Fig. 2)
is explained within the above
alternate procedure. The length dependence of the oscillation
frequency originates from the
smaller $(E_i - E_j)$ in longer chains.

\subsection{Dynamics in the PPP model}

We now present our results for interacting electrons in $H_{intra}$ and
$H_{inter}$. In all cases for the interchain $V_{i,j}$ we have
chosen the Ohno parameters, and the interchain hopping $t_{\perp}$ = 0.1 eV.
For $X_{\perp}$, we present the results of calculations with both
$X_{\perp}$ = 0 and 0.1 eV.  In Figs. 3 (a) and 3 (b)
we show the plots of $I_{mn}(t)$ in the singlet
and triplet channels for pairs of ethylenes, butadienes and 
hexatrienes, respectively, for the case of $X_{\perp}$ = 0. The same
results are shown in Figs 3 (c) and 3(d)
for $X_{\perp}$ = 0.1 eV.

The most obvious difference from the H\"uckel model is that the yields 
$I_{mn}(t)$ in both the
singlet and triplet channels are considerably reduced in the present cases.
Two other points are to be noted. First, there is now substantial 
difference between the singlet and triplet channels, with the singlet yield
higher in all cases. Second, the strong differences in singlet and triplet
yields are true for both $X_{\perp}$ = 0 and $X_{\perp} \neq$ 0. This is
in contradiction to the Golden Rule approach \cite{Shuai00a,Ye02a}, which
ignores the energy difference between the 1$^1B_u$ and the 1$^3B_u$.
The only consequence of nonzero $X_{\perp}$ is the
asymmetry between the yields of
$(1^1A_g)_1(1^1B_u)_2$ and $(1^1A_g)_2(1^1B_u)_1$ in the singlet channels, 
and a similar asymmetry
in the triplet channels. Further discussion of this asymmetry can be found 
in Appendix 1.
The overall conclusion that emerges from the results of 
Figs. 3 (a) - (d) 
is that nonzero electron-electron interactions substantially
enhances $\eta$.

In order to understand the above results in further detail we have also 
carried out the dynamics calculation for pairs of ethylenes according to 
Eq.~\ref{yld_t}. 
As in the H\"uckel case these calculations yield the same results as the
more general method. Our results for the wavefunctions of the composite
two-chain system and the overlaps of the product eigenstates of the final
neutral molecules with these are shown in Tables III and IV. The degeneracies
\begin{table}
\caption[]{Significant $c_i = <\Psi(0)|\psi_i >$ and the
$<m,n|\psi_i > $ values and
their product for PPP model in the
absence of electric field, for a pair of ethylenes in the singlet channel. 
The index $i$ corresponds to the index of `significant' eigenstates of the 
total system and $E_i$ the corresponding energy eigenvalue. }
\label{tbl_cff_ppp_s} \begin{center} \begin{tabular}{|c|c|c|c|c|} \hline
~~~~ & $E_i$ & $c_i$ & $< m,n|\psi_i > $ &  $c_i < m,n|\psi_i > $ \\ \hline
4  & 0.5295 &  -0.0249 & -0.6992 &  .0174 \\
5  & 0.7328 &  -0.0458 & -0.6953 &  .0318 \\
11 & 3.7748 &   0.7066 & -0.0258 & -.0182 \\
13 & 3.7844 &  -0.7056 & 0.0446  & -.0315 \\
29 & 11.2503 &  0.0082 & 0.1020  &  .0008 \\
30 & 11.6379 & -0.0025 & 0.1206  & -.0003 \\
32 & 14.0483 &  0.0050 & 0.0028  &  .00001 \\
34 & 14.0611 & -0.0054 & 0.0081  & -.00004 \\
\hline
\end{tabular}
\end{center}
\end{table}
\begin{table}
\caption[]{Significant $c_i = < \Psi(0)|\psi_i > $ and the
$< m,n|\psi_i > $ values and their
product for PPP model in the absence of electric field, for a pair of 
ethylenes in the triplet channel.}
\label{tbl_cff_ppp_t} 
\begin{center} 
\begin{tabular}{|c|c|c|c|c|c|} \hline ~~~~
& $E_i$ & $c_i$ & $<mn,n|\psi_i >$ & $c_i < m,n|\psi_i > $  \\  \hline
2 & -2.7283 &  -0.0215  & -0.6980 &  .0150 \\
3 & -2.7238 &  -0.0091  &  0.6982 & -.0064 \\
10 & 3.7697 &   0.7068  &  0.0060 &  .0042  \\
12 & 3.7775 &  -0.7067  &  0.0203 & -.0143  \\
19 & 8.0804 &  -0.0056  &  0.1115 & -.0006\\
20 & 8.0875 &  -0.0190  & -0.1114 &  .0021 \\
31 & 14.0475 &  0.0051  & -0.0056 & -.00003\\
33 & 14.0515 & -0.0052  & -0.0023  & .00001\\
\hline
\end{tabular} \\
\end{center}
\end{table}
in the eigenstates of the composite system that characterized the H\"uckel
model are now lifted, which is a known electron correlation effect. What is
more significant in the present case is that the composite state wavefunctions
that have large overlaps with $\Psi(0)$ are now not the same ones that have
large overlaps with the product wavefunctions of the final states. This is
what reduces the yields of the charge-transfer processes in the
PPP model, relative to the H\"uckel model. 

Tables III and IV give a clear physical picture of the charge recombination 
process. For a large
yield what appears to be essential is that {\it the composite two-chain
system must have at least some eigenstates which have simultaneously large 
overlaps with
both the direct product of the initial polaronic states and the direct product
of the pair of eigenstates of the neutral subsystems in the
the chosen channel.} This can be interpreted as a ``transition state theory''
for the charge recombination reaction of Eq.~\ref{CT}.
Large overlaps with the initial polaronic pair states occur for the states 11
and 13 in the singlet channel (see Table III), and for the states 10 and 12
in the triplet channel (see Table IV). 
This is in contrast to the H\"uckel case, where the
large overlaps with the polaron pair wavefunctions were with the same
composite two-chain eigenstates. The overlaps of these specific
two chain eigenstates are larger for
products of singlet final states $|1^1A_g\rangle_1|1^1B_u\rangle_2$
than for triplet final states $|1^1A_g\rangle_1|1^3B_u\rangle_2$, and this is
what gives a larger yield for the singlet exciton.

\subsection{Effects of external electric field}

Our results in the previous subsection already indicate that $\eta$ can be
substantially larger than 0.25 for the correlated electron Hamiltonian of 
Eq.~\ref{PPP}. From comparison of Fig. 2
and Figs. 3 (a) - (d), we see however, that the relative
yields $I_{mn}(t)$ are lower by orders of magnitude for interacting electrons.
This is easily understandable within {\it time-independent} second order
perturbation theory, within which the extent to which the initial 
polaron-pair state is modified is directly proportional to 
the matrix element of $H_{inter}$ between
the initial and final states, and inversely proportional to 
the zeroth order energy difference. Since the energy differences between
the polaron-pair states and the final neutral states are substantial within
the PPP Hamiltonian, the yields are low. 
There are two possible interpretations of these results. 
First, the actual yields
of excitons in OLEDS is indeed low, compared to the theoretical maximum for
noninteracting electrons (recall that no direct comparison of
the experimental light emission intensities with the theoretical maximum is
possible). Second, the experimentally observed yields are influenced 
substantially by external factors ignored so far. We consider this second
possibility here, and calculate within our time-dependent formalism the yields
$I_{mn}(t)$ in the presence of an external electric field (``external'' in
the following
includes the effects of both the actual bias voltage as well as all internal
field effects). What follows may be thought of as overly simple, but 
nevertheless, we believe that it gives the correct physical picture. We first
present our formalism and numerical calculations, and only then we discuss the
interpretation of these results. 

As before, we consider pairs of molecules that are parallel to each other,
with the molecular chain-axes aligned parallel to the x-axis. The electric
field is chosen along the y-axis, such that the total Hamiltonian now has an
additional contribution,
\begin{eqnarray}
H_{field} = E \sum_i ((n_i-1).y_i + (n_i^\prime-1).y_i^\prime)
\label{field}
\end{eqnarray} 
In the above $E$ is the strength of the electric field, and $y_i$ ($y_i^
\prime$) gives the y-component of the 
location of the i{\it th} (i$^{\prime}${\it th} 
carbon atom in molecule 1 (2). 
We now perform our
dynamical calculations with the complete Hamiltonian including $H_{field}$.

In Figs. 4 and 5
we show the effect of the external 
electric field on the yield
in the singlet and triplet channels for a pair of ethylenes. We see
that in all the cases there is a strong nonlinear
dependence of the yield on the external field. In both the singlet and the
triplet channels, we see sharp increases in the yields over a range of field
strengths. The field strengths at which the increases in the yields
occur are about two orders of magnitude larger than the experimental fields
in the OLEDS, and we comment upon this below.
Here we only observe that the
field strength $E$ over which the singlet yield is larger is smaller
than field strength over which the 
triplet yield dominates.

We have performed similar calculations for the longer chain systems, and in
all cases the effects are the same, viz., there 
exists a range of field strength
where a sudden increase in the singlet yield occurs, while at still larger
fields there occurs a similar jump in the triplet yield.
In Figs 6 (a) and 6 (b)
we have shown the singlet 
and triplet yields for field strengths
of 0.3 V/\AA~ and 1.0 V/\AA~, respectively, for hexatriene. 
In general, for a given spin channel the threshold field strength 
decreases with the chain length (the threshold field for the
singlet channel decreases from 0.7 V/\AA~ to 0.3 V/\AA~ on going from 
ethylene to hexatriene, while the threshold field for the triplet 
channel decreases from 1.6 V/\AA~ to 1.0 V\AA~). The most important 
conclusions that emerge from these calculations are that, (a) 
{\it macroscopically observable} yields, comparable to the
zero-field yields within the noninteracting H\"uckel model, are found for
large fields, and (b) while the calculated $\eta$ are greater than 
0.25 for smaller fields, 
this is reversed with further increase in the field strength.

In order to understand the origin of the increased yields over ranges
of the electric field, we have analyzed
the case of a pair of ethylenes extensively, within Eq.~\ref{yld_t}. 
Firstly it is worth noting that
the geometry in which the field is introduced, the product states of the
neutral Hamiltonian are unaffected by the electric field.  We also notice
that the eigenvalues of the total Hamiltonian are not very sensitive to the
external field.  As in the field-free cases, we have obtained the 
projections of the eigenstates of the two-chain system on the initial 
state as well as the product of the final states, as a function of the 
applied electric field in both the singlet and the triplet
channels. In Fig. 7,
we plot the coefficients
$\langle \psi_i|m,n \rangle$ as function of the electric field for
the singlet and the triplet channels. We see that there are several
states that show strong variation in both cases as a function of the field.
However,
when a product of these coefficients with $<\Psi(0)|\psi_i>$ is analyzed,
the number of the states that simultaneously have a large value of these
coefficients at the same electric field is smaller. In Fig. 8,
we plot the dominant coefficients
of these projections, as a function of the applied field. We note that only a
few states have both projections simultaneously large. We also note that both
the projections peak at the same field strength. It is this that leads to an
abrupt increase in the yield at that field strength.

The eigenstates of the full Hamiltonian that have large projections
simultaneously to both the initial and the final states can in fact
be expressed almost completely as a linear combination
of the initial polaron product state and the final product state of the
neutral system eigenstates. In Tables V and VI, we show the projections
\begin{table}
\caption[]{ In case of a pair of ethylenes the states with significant
$c_i=<\Psi(0)|\psi_i>$ and $<m,n|\psi_i>$, for the PPP
model
with electric field of 0.7 V/\AA ~ in the singlet channel. }
\label{tbl_eth_sing_ppp}
\begin{center}
\begin{tabular}{|c|c|c|c|c|} \hline
  i & $E_i$~(eV) & $c_i$ & $<m,n|\psi_i>$ &
$c_i<m,n|\psi_i>$ \\ \hline
4 & 0.5153 & 0.2086 & 0.7629 & 0.1591 \\
5 & 0.6828 & 0.3824 & 0.4933 & 0.1886 \\
11 & 1.0479 & -0.9001 & 0.3868 & -0.3482 \\
21 & 6.5753 & 0.0005 & -0.0055  & 0.0000 \\
27 & 11.2881 & -0.0016 & -0.0577 &  0.0001\\
30 & 11.6946 & 0.0040 & -0.1175 & -0.0005 \\
\hline
\end{tabular}
\end{center}
\end{table}
\begin{table}
\caption[]{ In case of a pair of ethylenes the states with significant
$c_i=<\Psi(0)|\psi_i>$ and $<m,n|\psi_i>$, for the PPP model
with electric field of 1.6 V/\AA ~ in the triplet channel. }
\label{tbl_eth_trip_ppp}
\begin{center}
\begin{tabular}{|c|c|c|c|c|} \hline
  i & $E_i$~(eV) & $c_i$ & $<m,n|\psi_i>$ &
$c_i<m,n|\psi_i>$  \\ \hline
2 & -2.8347 &  0.5927 & 0.3031 & 0.1796 \\
3 & -2.7237 &  0.0052 & 0.9142 & 0.0048 \\
5 & -2.5174 &  0.8053 & -0.2179 & -0.1755  \\
20 & 7.5928 &  0.0089  & -0.0180 & -0.0002\\
23 & 8.0821 &  0.0054  &  0.1518 & 0.0008\\
24 & 8.1387 & -0.0031 & 0.0364   & -0.0001\\
26 & 10.1769 & 0.0001 & 0.0163 & 0.0000 \\
\hline
\end{tabular}
\end{center}
\end{table}
of the eigenstates of the full Hamiltonian at the resonant electric field
on to (i) the initial state and
(ii) to the product of eigenstates of the neutral system for which resonance
is observed. We note that there are a few eigenstates of the full Hamiltonian
which have large coefficients for both projections. This seems to be
independent of the energy of the eigenstate of the total system. The energetics
decide the period of oscillations and not the amplitude of the oscillations. 

We now come to our interpretations of the above numerical calculations. 
In all cases
the applied fields in our calculations
are substantially larger than what is expected from the
externally applied voltage in OLEDS. Note, however, that our molecules are
rather small, and the calculated threshold fields at which the effect becomes
observable decrease with the molecular size. In this context, it is worth
recalling a previous exact calculation of electroabsorption for short finite
polyenes \cite{DGuo}. There the electric field was parallel to the chain axis
(as opposed to perpendicular, as in the present case), and it was found that
the calculated electroabsorption can simulate the experimentally observed
behavior in long chain polymers \cite{Weiser,Liess}, 
provided the electric field used in the short
chain calculation was larger by two orders of magnitude than the experimental
field. This is because of the large energy gaps in short chains.
We believe that a similar
argument applies in the present calculations of interchain charge-transfer:
the energy difference between the initial polaron-pair state and the final 
states is much larger in the small molecule-pair 
system than in the experimental
systems, even when oligomeric. The analogy to electroabsorption would then 
imply that the enhanced macroscopic yields would occur in the real systems at
much smaller, perhaps even realistic fields. 

One final point concerns the geometry used in our
calculations. In real OLEDS the relative orientations of the molecules of a 
given pair, as well as the orientation of the electric field with respect to
individual members of the pair will both be different from that assumed in our
simple calculations above. Electric fields that are nonorthogonal to the 
chain axis of a molecule will have even stronger effects than found in our
calculations \cite{DGuo}, while the random arrangements of the molecule pairs
with respect to the field in the experimental systems
implies that the range of field over which a given
spin channel dominates will be substantially larger than that found in our 
calculations. We therefore believe that a proper interpretation of our 
calculations is that in the experimental systems, there occur macroscopically
large yields of both singlet and triplet excitons over a broad range of
electric field. For small to moderate field strengths, the singlet
channel dominates over the triplet channel. However, at still larger fields
it is possible that this situation reverses. Whether or not this higher regime
of field strength is experimentally accessible is a topic of future 
theoretical and experimental research.

\subsection{Chain length dependence}

We now discuss the chain length dependence of $\eta$ as has recently been
determined experimentally \cite{Wilson01a,Wohlgenannt02a}. From careful
measurements using different techniques, Wilson et. al.\cite{Wilson01a} 
and Wohlgenannt et. al.\cite{Wohlgenannt02a} have established that
$\eta$ increases with conjugation length. Wilson et. al. have shown
that while $\eta$ is close to the statistically expected 0.25 in the monomer
\cite{Wilson01a}, it is substantially larger in the polymer. 
Wohlgenannt et. al. have
shown that $\sigma_T/\sigma_S$ increases linearly with the inverse of the
conjugation length.

Within our numerical procedure, it is difficult to determine the chain
length dependence of $\eta$ directly. This is because of multiple reasons,
which include, (i) the limitation to rather small sizes, (ii) the necessity
to integrate $I_{mn}(t)$ over a complete period in each case
in Figs. 3 (a) - (d) to obtain the total yield over that period, 
and (iii) the 
difference in the periods for singlet and triplet channels, as well as the
differences among different singlet and triplet channels. We therefore present
our discussion of chain length dependence within a simplified formalism that
is consistent with our time-dependent procedure.

Consider a transition (which could be of the charge-transfer type) between
the states $|k\rangle$ and $|s\rangle$ of a two-state system, such 
that at time $t$ = 0
the system is in state $|s\rangle$ ($c_s(0)$ = 1, $c_k(0)$ = 0). 
We are interested in the yield $|c_k(t)|^2$ at a later time $t$ due to a
perturbation $V_{ks}$. This is a standard textbook problem \cite{Merzbacher},
and the time-dependent Schrodinger equation in this case is, 
\begin{equation}
\label{twolevel}
i\hbar{\partial\over\partial t}c_k = V_{ks}exp(i\omega_{ks}t)
\end{equation}
For time-independent $V_{ks}$, as is true here ($V_{ks}$ in the present case
is simply $H_{inter}$ of Eq.~\ref{inter}) the above equation is easily 
integrated to give,
\begin{equation}
\label{integrated}
|c_k(t)|^2 = 2|\langle k|V|s\rangle|^2 {{1 - \cos \omega_{ks}t}\over{(E_k^{(0)}
- E_s^{(0)})^2}}
\end{equation}
In our case $|s\rangle = |P^+_1P^-_2\rangle$, with the appropriate spin
combinations, and $|k\rangle = |(1^1A_g)_1(1^1B_u)_2\rangle$ for the
singlet channel, and  $|k\rangle = |(1^1A_g)_1(1^3B_u)_2\rangle$ for the
triplet channel (as usual, $|k\rangle$ can also have the chain indices
reversed). We have already demonstrated (see section III and Appendix 1) that
the matrix element $\langle k|V|s\rangle$ is nearly the same for the singlet
and triplet channels, except near the unique point $X_{\perp}/t_{\perp}$ = 
0.5. Ignoring
the oscillation involving $\omega_{ks}$ we note that the relative yield
of the singlet exciton is inversely proportional to the square of
$\Delta E_S = [E(P^+) + E(P^-) - E(1^1A_g) - E(1^1B_u)]$, while that of the
triplet exciton is  inversely proportional to the square of
$\Delta E_T$, in which $E(1^1B_u)$ in the above
is replaced with $E(1^3B_u)$ (note, however,
that within the two-state approximation 
we have {\it assumed} that the singlet and triplet states
that are of interest are the lowest singlet and triplet states; this can
only be justified by the complete many-state calculations of the previous
subsections).
We see 
immediately that this simple two-state formalism predicts higher
singlet yield, since $E(1^1B_u)$ is considerably higher than 
$E(1^3B_u)$. Importantly, the chain-length dependence of $\eta$ is also
understood from the above. Both $\Delta E_S$ and $\Delta E_T$ decrease with
increasing chain length. However, the ratio $\Delta E_S/\Delta E_T$
also decreases, because of the covalent character of the 1$^3B_u$ and
the ionic character of the 1$^1B_u$.
This is seen most
easily in the limit of the simple Hubbard model for the individual
chains (zero intersite Coulomb interaction and zero bond alternation in
Eq.~\ref{PPP}), where $\Delta E_S$ approaches 0 and $\Delta E_T$ approaches
$U$ in the long chain limit. 

We have calculated $\Delta E_S$ and $\Delta E_T$ exactly for all chain 
lengths N = 4 - 10 within the PPP-Ohno potential. While the ratio 
$\Delta E_T/\Delta E_S$ shows the correct qualitative
trend 
(viz., increasing $\Delta E_T/\Delta E_S$
with increasing N) necessary for increasing $\eta$ with increasing N, 
the actual variation is small. This is to be expected,
since our chain length variation is small, and the Ohno potential decays
very slowly. With our limitation on N, it is necessary that the Coulomb
potential is short range, such that we have the same Hamiltonian at all chain
lengths, as is approximately true for the experimental systems investigated
\cite{Wilson01a,Wohlgenannt02a}. We have therefore done exact calculations of
$\Delta E_S$ and $\Delta E_T$ for the extended Hubbard Hamiltonian
($V_{ij}$ in Eq.~\ref{PPP} limited to nearest neighbor interaction $V$)
with parameters $t_{ij}$ = 1.08$t_0$ and 0.92$t_0$ for double and single bonds,
$V/t_0$ = 2 and $U/t_0$ = 5 and 6. In Fig. 9
we show our calculated results
for $\Delta E_T/\Delta E_S$ for the two cases, for different N. In both
cases, increasing $\Delta E_T/\Delta E_S$ with increasing N indicates larger
$\eta$ for longer chain lengths. Energy convergences are faster with larger
$U$, which explains the steeper behavior of $\Delta E_T/\Delta E_S$ 
for larger $U$, and gives additional support to our argument.

\subsection{Role of heteroatoms}

The experiments by Baldo et. al. \cite{Baldo99a} and Wilson et. al. 
\cite{Wilson01a} both indicate that in small molecular systems $\eta$ can
be close to 0.25. This is in contrast to our results for ethylene (see Fig. 3).
for which $\eta$ is calculated to be substantially larger. One reason for 
this might be that the Coulomb correlation effects in thin film samples
are smaller than within the PPP Hamiltonian due to intermolecular interactions.
The dominant effect, however, is due to the heteroatoms in the
molecules investigated by these authors, as we show below. Specifically,
the site-energy (electronegativity) difference between the heteroatom and
carbon atoms makes these systems closer to the H\"uckel limit and this is
what decreases $\eta$.

In order to compare with the model polyene systems we consider pairs of
(CH=N)$_2$ in the following calculations. The single chain Hamiltonian 
(Eq.~\ref{PPP}) is then
modified as follows \cite{Albert91a}. 
The Hubbard $U$ for the nitrogen atoms, $U_N$ = 12.34. eV.
The local chemical potential $z_N$ for nitrogen with $\pi$ lone pair involved
in conjugation is 2. Finally, nitrogen has
site energy $\epsilon$ =  -- 2.96 eV relative to that of the carbon atoms. 
There are
two possible arrangements for the two chains in a parallel configuration, 
(i) carbon (nitrogen) on one chain lying directly above carbon (nitrogen)
atom on the other, and (ii) carbon atom on one chain lying above nitrogen
atom on the second chain. We have chosen arrangement (i), -- there is no
fundamental reason for arrangement (ii) to have a very different $\eta$.

In Figs. 10 (a) and (b) we have plotted the $I_{mn}(t)$ 
for the singlet and triplet channels, respectively, 
for the case of $X_{\perp}$ = 0.
Figs. 10 (c) and (d) show the same for $X_{\perp}$ = 0.1 eV. 
The most important conclusion that emerges from these calculations is that
the relative yields of triplets are substantially larger in the present case,
so much so that $\eta$ can be even close to the statistical limit of 0.25
(note that there are three triplet channels and the figures show the results
for only one of these). We believe that these results give a qualitative
explanation of the observation of Baldo et. al. 
\cite{Baldo99a}. Taken together
with the chain length dependence of $\eta$, as found in the previous 
subsection, these results also explain qualitatively the observations by
Wilson et. al. \cite{Wilson01a}, since the same chain length dependence
found in the case of simple polyenes should be true here also, although
it is conceivable that rate of increase of $\eta$ with N here may be slower.

\section{Discussions and conclusions}
\label{conclusion}

With a parallel arrangement of two polyene chains, we have shown that
several experimentally observed
qualitative features of the singlet-to-triplet yield ratios in 
$\pi$-conjugated systems can be understood within a well-defined total
Hamiltonian for the two-chain system. While our model systems are rather
simple, our theoretical treatment of the charge-transfer process between
the two chains is exact. We have given a full time-dependent approach to
the interchain charge-transfer process, and have shown that in systems
containing only carbon atoms, the overall yield of the singlet exciton
is considerably larger than that of triplet excitons and $\eta >$ 0.25.
This is a direct consequence of moderate electron-electron Coulomb 
interactions which has strong effects on both the energies and the 
wavefunctions of the singlet and triplet excitons. 
The mechanism of the exciton yields that emerges from our calculations is
as follows. For large yields, it is essential that there exist excited
states of the composite two-chain system whose wavefunctions have 
simultaneously
large overlaps with the wavefunction of the
initial state consisting of polaron pairs, and the final state consisting
of the two chains in the neutral states. Overlaps of the excited states
of the two-chain system with final states in the singlet channel are
considerably larger than for final states in the triplet channel, and this
is what gives a large yield for the singlet exciton. This is a consequence
of the different natures of the singlet and triplet excitons, which are
ionic and covalent, respectively, in the VB notation.
Our result here is consistent with experiments on long oligomers and
polymers \cite{Cao99a,Ho00a,Wohlgenannt01a,Wilson01a,Wohlgenannt02a}.
Although our exact calculations are limited to short chains, within a
two-state approximation that is consistent with the full multi-level
calculation we have shown that $\eta$ increases with the chain length,
in agreement with experimental observations \cite{Wilson01a,Wohlgenannt02a}.
The two-state approximation gives an alternate explanation of the higher
yield of the singlet exciton that is related to the singlet and triplet
exciton binding energies, which are substantially different in 
$\pi$-conjugated polymers.  Finally, we have examined the 
role of heteroatoms, and have shown that in small
molecular systems with nitrogen as the heteroatom, $\eta$ is substantially
smaller, and may be even close to the statistically expected value of 0.25.
The wavefunctions in this case, due to the strong electronegativity difference
between the heteroatom and carbon atoms, are closer to the H\"uckel limit,
and this is what increases the relative yield of the triplet exciton. Our
results here successfully explain the difference between Alq$_3$\cite{Baldo99a}
and heteroatom
containing monomers \cite{Wilson01a} on the one hand, 
and polymeric systems on the other, and thereby provide additional
strong support to our theoretical approach.

The time-dependent approach to the charge-transfer process developed here
is completely general and can be applied to many other similar
processes, for example, photoinduced charge-transfer, triplet-triplet 
collisions in OLEDS, etc. These and other applications are currently being
investigated. Similarly, for a more complete understanding of the
chain length dependence of $\eta$, we will investigate charge-transfer
process within the density matrix renormalization group technique.

While this manuscript was under preparation we received a manuscript 
(S. Karabunarliev and E.R. Bittner, cond-mat/0206015) from E. Bittner
that discusses the relative yields of singlet
and triplet excitons within the context of intrachain processes (see
section ~\ref{theories}) as opposed to the interchain process discussed
here. Although the approach of these authors is different from ours, they
also find that the relative yields of singlet and triplet excitons
are determined by their binding energies (smaller binding energies giving
larger yields). It is not clear whether the approach used by these authors
applies to molecule-based OLEDS.
These authors have also investigated the effect of
broken electron-hole symmetry, which is related to our calculations on
chains of (CH=N)$_2$. Our results here are different. While Karabunarliev
and Bittner find even higher relative yield of singlet excitons (compared
to electron-hole symmetric case) we find that $\eta$ here is smaller
(see above). While a complete analysis of the electron-hole recombination
must include both interchain and intrachain processes (and is a subject of
future work in this area), we believe that 
this last result, when compared to experiments, justify our basic assumption
that spin-dependence of the yields of excitons can be understood largely
within the context of intermolecular and interchain charge-transfer.

\section{Acknowledgments}

Work in Bangalore was supported by the CSIR, India and DST, India,
through /INT/US
(NSF-RP078)/2001. Work in Arizona was supported
by NSF-DMR-0101659 and NSF-INT-0138051. We are grateful to our experimental
colleagues Z.V. Vardeny and M. Wohlgenannt for numerous stimulating
discussions. S.M. acknowledges the hospitality of the Indian Institute of
Science, Bangalore, where this work was completed.
 
\section{Appendix 1}
\label{appendix}
We present here detailed analytic calculations of the matrix elements of
$H_{inter}$ for the case of two ethylenes. We believe that these calculations 
give clear understandings of the chain-length independence of the calculated 
$\sigma_S/\sigma_T$ within the Fermi Golden Rule approach (Eq.~\ref{fermi})
that was presented 
in section \ref{theories} (see Fig. 1).
We also believe that
even as these calculations show the inadequacy of the Golden Rule approach they
provide an indirect understanding of the actual mechanism
behind large $\eta$ in long chain polymers.

As in the rest of the paper we consider parallel arrangements of the ethylene
molecules, with sites 1 and 2 (3 and 4) corresponding to the lower (upper)
molecule. Subscripts 1 and 2 that are assigned to wavefunctions describe
the lower and upper molecule, respectively.
The relevant single-molecule eigenstates, corresponding to the
lower molecule then can be written as
\begin{eqnarray}
|1^1A_g\rangle_1 = (c_1/\sqrt{2})(a^{\dagger}_{1,\uparrow}a^{\dagger}_
{1,\downarrow} + a^{\dagger}_{2,\uparrow}a^{\dagger}_
{2,\downarrow})|0\rangle~ +~
~~~~~~~~~~~~~~~~~~~~\nonumber \\
(c_2/\sqrt{2})(a^{\dagger}_{1,\uparrow}a^{\dagger}_{2,\downarrow} -
a^{\dagger}_{1,\downarrow}a^{\dagger}_{2,\uparrow})|0\rangle
~~~~~~~~(16a)~~~~~~~~~~~\nonumber \\
|1^1B_u\rangle_1 = (1/\sqrt{2})(a^{\dagger}_{1,\uparrow}a^{\dagger}_
{1,\downarrow} - a^{\dagger}_{2,\uparrow}a^{\dagger}_
{2,\downarrow})|0\rangle
~~~~~~~~~(16b)~~~~~~~~~~~\nonumber \\
|1^3B_u\rangle_1 = (1/\sqrt{2})(a^{\dagger}_{1,\uparrow}a^{\dagger}_
{2,\downarrow} +
a^{\dagger}_{1,\downarrow}a^{\dagger}_{2,\uparrow})|0\rangle 
~~~~~~~~~(16c)~~~~~~~~~~~\nonumber \\
|P^+\rangle_1 = (1/\sqrt{2})(a^{\dagger}_{1,\sigma} + 
a^{\dagger}_{2,\sigma})|0\rangle
~~~~~~~~~~~~~~~~~~~~(16d)~~~~~~~~~~~\nonumber \\
|P^-\rangle_1 = (1/\sqrt{2})(a^{\dagger}_{1,\uparrow}
a^{\dagger}_{1,\downarrow}a^{\dagger}_{2,\sigma} - 
a^{\dagger}_{2,\uparrow}a^{\dagger}_{2,\downarrow}a^{\dagger}_{1,\sigma})
|0\rangle 
~(16e)~~~~~~~~~~~\nonumber 
\label{states1}
\end{eqnarray}
In the above $|0\rangle$ is the vacuum for chain 1 and $c_1$ and $c_2$ are 
the coefficients of the ionic and covalent configurations in the 1$A_g$ 
ground state that are to be determined by solving for the 2 $\times$ 2 
$A_g$ subspace of ethylene within the PPP Hamiltonian ($c_1$ = $c_2$ = 
1/$\sqrt{2}$ in the H\"uckel Hamiltonian and the matrix elements in 
Eq. ~\ref{fermi} in the singlet and triplet channels
are exactly equal in this case). 
We have chosen the $M_S$ = 0 wavefunction for the $1^3B_u$, but what follows 
is equally true for the $M_S = \pm$ 1 wavefunctions. We have not assigned 
definite spin states to the charged polaronic wavefunctions, since the 
charged molecules can have either spin, and since different
combinations of these spin states give the initial spin singlet or triplet
product eigenstates for $|P^+P^-\rangle$. Note, however, the relative
minus signs between the two configurations in $|P^-\rangle$, as
opposed to the relative plus signs between the two configurations in 
$|P^+\rangle$. This is what ensures that the product wavefunctions of the type
$|P^+_1P^-_2\rangle$, with positive charge on molecule 1 and negative charge 
on molecule 2 has odd parity with respect to the center of inversion on a 
single chain, and charge recombination can therefore only generate neutral 
states that have odd parity (for example, $|1^1A_g\rangle_1|1^1B_u\rangle_2$ 
but not $|1^1A_g\rangle_1|2^1A_g\rangle_2$).

We consider the initial states $|i_S\rangle$ and $|i_T\rangle$
first, which are constructed from taking products of the polaronic 
wavefunctions given above. Since these product functions contain four terms
each, and also since one of our goals is to arrive at a visual representation
of the charge recombination process in configuration space, we have chosen
not to write their explicit form but have given in Fig. 11
the wavefunctions in the VB notation, where
a singlet bond between sites $i$ and $j$ is defined as
2$^{-1/2}(a^{\dagger}_{i,\uparrow}a{\dagger}_{j,\downarrow} -
a^{\dagger}_{i,\downarrow}a{\dagger}_{j,\uparrow})|0\rangle$, a
triplet bond (with an arrow pointing from site $i$ to site $j$) is defined as
2$^{-1/2}(a^{\dagger}_{i,\uparrow}a{\dagger}_{j,\downarrow} +
a^{\dagger}_{i,\downarrow}a{\dagger}_{j,\uparrow})|0\rangle$, and crosses
correspond to doubly occupied sites 
$a^{\dagger}_{i,\uparrow}a{\dagger}_{j,\downarrow}|0\rangle$.  
Given the initial and final states, it is now easily seen that 
$V_{i,j}$ in $H_{inter}$ (Eq.~\ref{inter}) plays no role within the 
Golden Rule approach \cite{Shuai00a,Ye02a}, since this
term causes no transition between the initial and final states (note, 
however, $V_{i,j}$ can play a significant role in the full dynamics
calculation of section \ref{prop}). The matrix elements of the remaining 
terms in $H_{inter}$ are now readily evaluated and these are given below,
\begin{eqnarray}
\langle (1^1A_g)_1(1^1B_u)_2|H_{inter}|P^+_1P^-_2\rangle_S 
~~~~~~~~~~~~~~~~~~~~~~~~~~~~~~~ \nonumber\\
= - (c_1/\sqrt{2})(-t_{\perp} + 2X_{\perp}) - (c_2/\sqrt{2})(-t_{\perp} 
+ X_{\perp}) ~~ (17a)\nonumber \\ 
\nonumber \\
\langle (1^1A_g)_2(1^1B_u)_1|H_{inter}|P^+_1P^-_2\rangle_S 
~~~~~~~~~~~~~~~~~~~~~~~~~~~~~~~\nonumber \\
= (c_1/\sqrt{2})(-t_{\perp} + 2X_{\perp}) + (c_2/\sqrt{2})(-t_{\perp} + 
3X_{\perp}) ~~~(17b) \nonumber \\ 
\nonumber \\
\langle (1^1A_g)_1(1^3B_u)_2|H_{inter}|P^+_1P^-_2\rangle_T 
~~~~~~~~~~~~~~~~~~~~~~~~~~~~~~~\nonumber \\
= - (c_1/\sqrt{2})(-t_{\perp} + X_{\perp}) - (c_2/\sqrt{2})(-t_{\perp} + 
2X_{\perp}) ~~(17c) \nonumber \\ 
\nonumber \\
\langle (1^1A_g)_1(1^3B_u)_2|H_{inter}|P^+_1P^-_2\rangle_T 
~~~~~~~~~~~~~~~~~~~~~~~~~~~~~~~\nonumber \\
= (c_1/\sqrt{2})(-t_{\perp} + 3X_{\perp}) + (c_2/\sqrt{2})(-t_{\perp} + 
2X_{\perp})~~~(17d) \nonumber 
\label{mat_elements}
\end{eqnarray}

\noindent Several points are to be noted now. First, for $X_{\perp}$ = 0, the 
squares of all
the matrix elements are equal, and hence there is no difference between
singlet and triplet generation within the Golden Rule approach in this limit,
and we agree on this with Shuai et. al. \cite{Shuai00a} 
Second, however, defining overall $\sigma_S$ as the sum of the squares 
of the matrix elements in Eqs. 17(a) and and (b), and $\sigma_T$ as 
the sum of the squares of the matrix elements in Eqs.  17(c) and (d), 
respectively, we see that $\sigma_S/\sigma_T$ depends very weakly on 
$X_{\perp}/t_{\perp}$ at
all $X_{\perp}/t_{\perp}$ except for $X_{\perp}/t_{\perp}$ very close to 0.5,
where ${-t_{\perp} + 2X_{\perp}}$ = 0 and ${-t_{\perp} + X_{\perp}}$ 
and ${-t_{\perp} + 3X_{\perp}}$ have opposite signs.
This is 
particularly so for the calculated $c_1$ and $c_2$ for PPP-Ohno parameters
($c_1$ = 0.5786, $c_2$ = 0.8156).
We now examine the different terms in Eqs.~17(a) - (d) in detail. From Fig. 11
we note that there are three classes of interchain electron
transfers: (i) charge-transfer between sites that are both singly occupied,
leading to a doubly occupied site or an empty site (denoted by
1 + 1 $\to$ 2 + 0, where the numbers denote site occupancies)
-- or the exact reverse
process, (ii) charge-transfers of the type 1 + 0 $\to$ 0 + 1, using the same
notation, and (iii) charge-transfers of the type 2 + 1 $\to$ 1 + 2, again
with the same notation. These three processes have different matrix 
elements $(-t_{\perp} + 2X_{\perp})$, $(-t_{\perp} + X_{\perp})$ and 
$(-t_{\perp} + 3X_{\perp})$, respectively. 
The role of $X_{\perp}$ now becomes absolutely clear. Nonzero $X_{\perp}$
creates an asymmetry between the upper and lower molecule, 
leading to a 
difference between the yields of
$|1^1A_g\rangle_1|1^1B_u\rangle_2$ and $|1^1A_g\rangle_2|1^1B_u\rangle_1$,
but it does not create a significant difference between $\sigma_S$ and
$\sigma_T$. 

At exactly $X_{\perp}/t_{\perp}$ = 0.5 terms containing 
$(-t_{\perp} + 2X_{\perp})$ in
the matrix elements vanish, while the other terms are also small and 
of opposite signs. The singlet
channel matrix elements now involve only $c_2$, while the triplet channel
matrix elements involve only $c_1$. Since for repulsive Coulomb interactions
$c_2 > c_1$, the sum of the the squares of the matrix elements here are
larger for the singlet channel than for the triplet channel. This is what
is reflected in our plot of Fig.~\ref{fermi}. Note, however, that the 
calculated yields approach zero in both cases here.
It is also clear from Eqs.~17 that this difference
between the singlet and triplet channels persist over a very narrow
region about $X_{\perp}/t_{\perp}$ = 0.5. We therefore do not believe that 
this is of any relevance for realistic systems.

Our final point concerns the chain-length independence of our results in
Fig.~\ref{fermi} (except near $X_{\perp}/t_{\perp}$ = 0.5). For arbitrary chain
lengths there can occur only the three classes of interchain charge-transfers
discussed above (1 + 1 $\to$ 2 + 0, 1 + 0 $\to$ 0 + 1 and 2 + 1 $\to$ 1 + 2). 
The detailed wavefunctions of longer chains are
different, but the expectation values 
$\langle n_{i,\uparrow}n_{i,\downarrow}\rangle$
for the different wavefunctions are nearly the same for fixed intrachain
correlation parameters. Thus although in long chains there can in principle
occur many more interchain hops that are of the type 1 + 1 $\to$ 2 + 0, such
charge-transfers lead to {\it additional} double occupancies (relative to the
overall initial states) that are energetically costly because of 
electron correlation effects. Such charge-transfers therefore make weak
contributions to the overall interchain charge-transfer.
The net consequence is the weak chain-length
dependence found in Fig. 1
at all points other than $X_{\perp}/t_{\perp}$ = 0.5. 

We believe that the above detailed calculation, aside from indicating the
inapplicability of the Golden Rule, also indicates that the proper theoretical
treatment must include the differences in the energies and wavefunctions of the
final states, as indeed is done in our time-dependent calculations.

\section{Appendix 2}
\label{appendix2}
Details of the numerical procedure that were not discussed in
section ~\ref{results} are given below.
The charged as well as neutral eigenstates of $H_{intra}$ for
individual chains are obtained in the VB representation by using a 
diagrammatic VB approach \cite{Ramasesha84a} with bit 
representation of the basis states. The eigenstates of a given spin
$S$ are obtained for 
$M_S=S$. The VB eigenstates are then transformed to the basis of
Slater determinants with $M_S=S$ by expanding the terms in each 
singlet pair and assigning an up-spin
at each unpaired site with single occupancy. Thus, a triplet VB
basis consisting of two singlet pairs and corresponding to a 
function with $M_S=1$ is expanded into four Slater determinants each
with $M_S=1$. To obtain eigenstate corresponding to other $M_S$ 
values with Slater determinantal basis, 
we apply the $\hat{S}^-$ operator on the state, as many
times as is necessary. 

We use the eigenstates of $|P^+\rangle$ and $|P^-\rangle$ 
to form the initial state of chosen spin in the form,
\begin{eqnarray}
    \Psi_{1,0}(0)  = {1 \over \sqrt{2}} \Bigg[|{{1} \over {2}}, +{{1} 
\over {2}}\rangle_1 
\star
|{{1} \over {2}}, -{{1} \over {2}}\rangle_2 \pm \\ \nonumber
|{{1} \over {2}}, -{{1} 
\over {2}}\rangle_1 \star |{{1} \over {2}}, +{{1} \over {2}}\rangle_2 \Bigg]
\end{eqnarray}
\noindent where the subscripts 1 and 0 refer to the total spin of the
initial state. The direct product of the states are expressed in the
Slater determinantal basis of the composite system with the coefficient
$c_k$ of the basis state $|k\rangle$ in the composite system being given by
\begin{equation}
    c_k = \sum_{l,m} d_ld_m <k|l \star m>
\end{equation}
\noindent where $d_l$ and $d_m$ are the coefficients of the basis states
$|l\rangle$ and $|m\rangle$ in the ground states of the subsystems 1 and 2 
respectively. 
The direct product itself is effected by shifting the
$2n_1$ bits of the integer representing the basis state of system 1 
with $n_1$ sites to the immediate left of the $2n_2$ bits in the integer 
that represents the basis state of system 2 with $n_2$ sites. The resulting 
larger integer with $2(n_1+n_2)$ bits correspond to an integer that 
represents one of the basis states of the composite system of $(n_1+n_2)$
sites.

The evolution of the initial state involves solving the linear algebraic
equations, ${\bf \rm A}{\bf x}(t+\Delta t) = {\bf b}$,
\noindent where, the matrix elements of the matrix {\bf \rm A} and 
the components of ${\bf b}$ are given by,
\begin{eqnarray}
     A_{ij} &=& (\delta_{ij} + i \hbar H_{ij} {{\Delta t} \over {2}}) \\
    b_i &=& \sum_j (\delta_{ij} - i \hbar H_{ij} {{\Delta t} 
\over {2}}) x_j(t) 
\end{eqnarray}
The matrix {\bf \rm A} in the largest system we have studied is nearly
of order one million and for reasonable convergence of the solution
of the system of equations we need a $\Delta t$ of the order of 
$0.05eV/{\hbar}$
which is typically 0.033 femtoseconds, and this guarantees 
diagonal dominance of the
matrix $\bf {\rm A}$. Thus, if one wishes to follow the dynamics 
for even as long as say 60 femtoseconds
one needs to solve the linear system about
2000 times. This is rendered possible by the sparseness of the matrix
$\bf {\rm A}$. For efficient convergence, we use a small matrix algorithm
\cite{Rett82a} which is very similar to the Davidson's algorithm 
for matrix eigenvalue
problem. In the case of the largest system size, it takes about twelve
hours on a DEC Alpha 333 MHz system to evolve the state by 60fs for a given
channel.

At the end of each iteration in the evolution of the state, we obtain the 
intensity or the yield in a pair of states in the neutral subsystem. The
number of such pairs is enormous when we deal with say two systems of
six sites each. The number of pairs in the singlet channel is 30,625
while that in the triplet channel is 33,075. We can reduce the pairs
onto which we project the evolved state by restricting ourselves
to a few low-energy states of the neutral subsystem. However, even in
this case, the number of pairs could be rather large. To overcome this
problem, we select only those pairs which have a minimum yield of say
$10^{-4}$ at all times. This restriction when implemented judiciously
leaves us with only a few pairs with significant yields.   

\bibliography{co}

\pagebreak

{\bf Figure Captions}

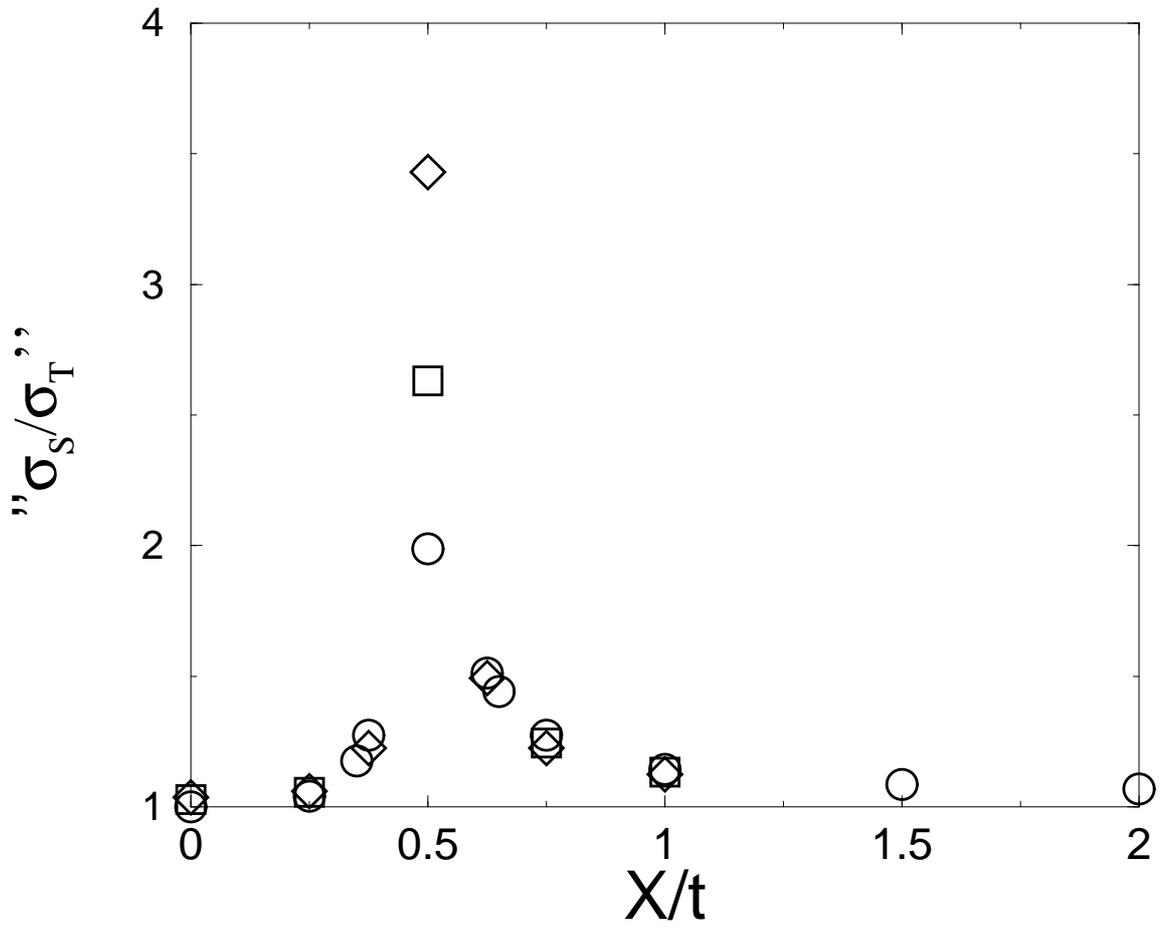
{\bf Figure 1 :} The ratio of the squares of the singlet and triplet matrix elements of
H$_{inter}$ ($\sigma_S/\sigma_T$ according to Eq. ~\ref{fermi}), as a
function of $X_{\perp}/t_{\perp}$ for pairs of ethylenes (circles),
butadienes (squares) and hexatrienes (diamonds).

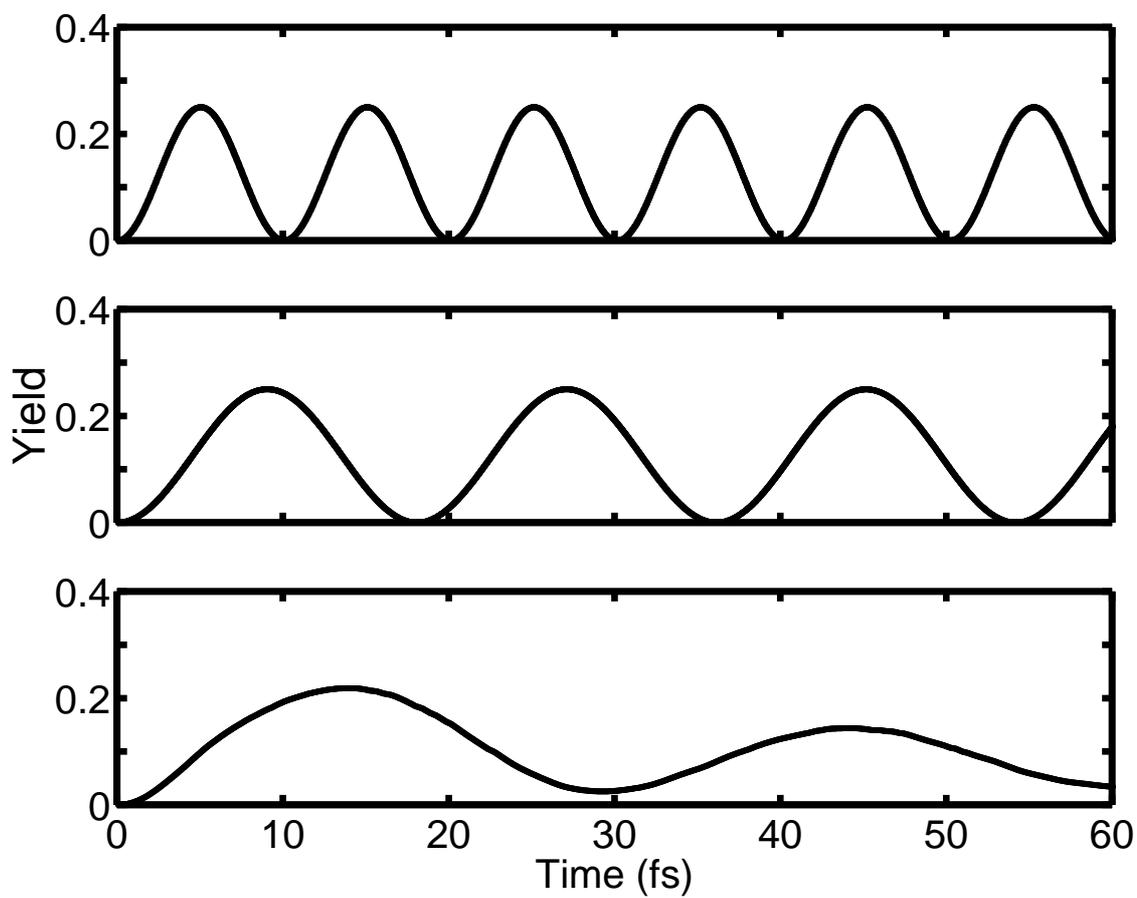
{\bf Figure 2 :} Yield in the singlet channel as a function of time,
for pairs of ethylenes (top panel), butadienes (middle panel)
and hexatrienes (bottom panel),within the simple H\"uckel model
($U=V_{ij}=X_{\perp}=0$).  Significant yield in all cases occur
only for final states $|(1^1A_g^+)_1(1^1B_u^-)_2>$ and
$|(1^1B_u^-)_1(1^1A_g^+)_2>$, between which the yields are identical.
Yields in the triplet channel  $|(1^1A_g^+)_1(1^3B_u^+)_2>$
and $|(1^3B_u^+)_1(1^1A_g^+)_2>$ are identical to those in the
singlet channel.

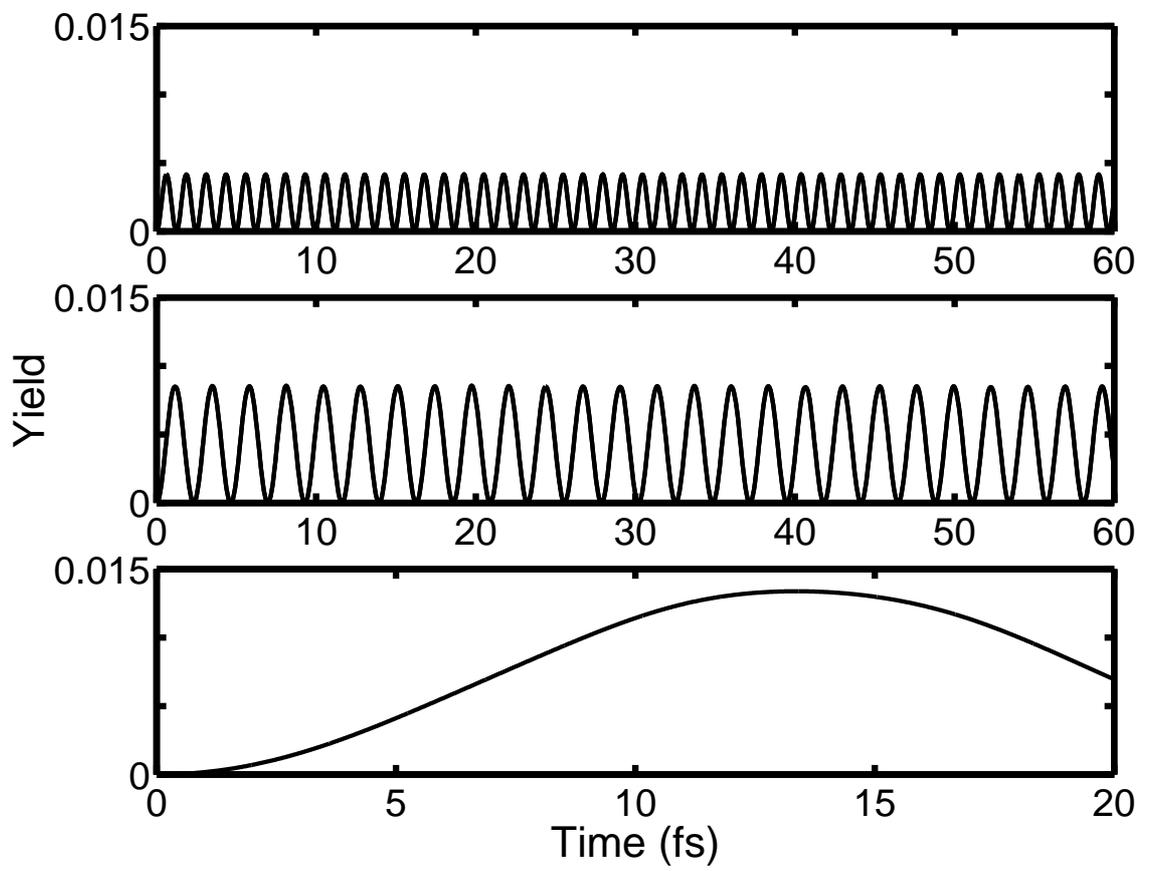
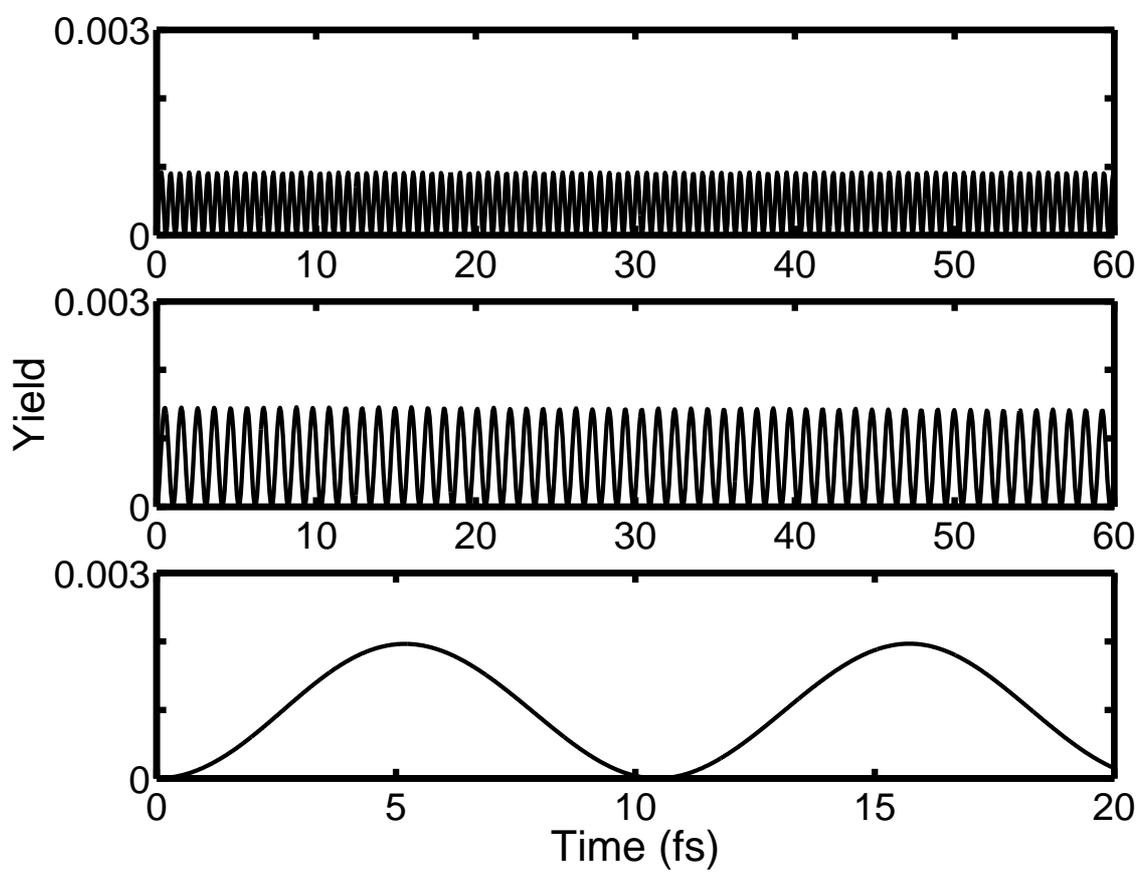
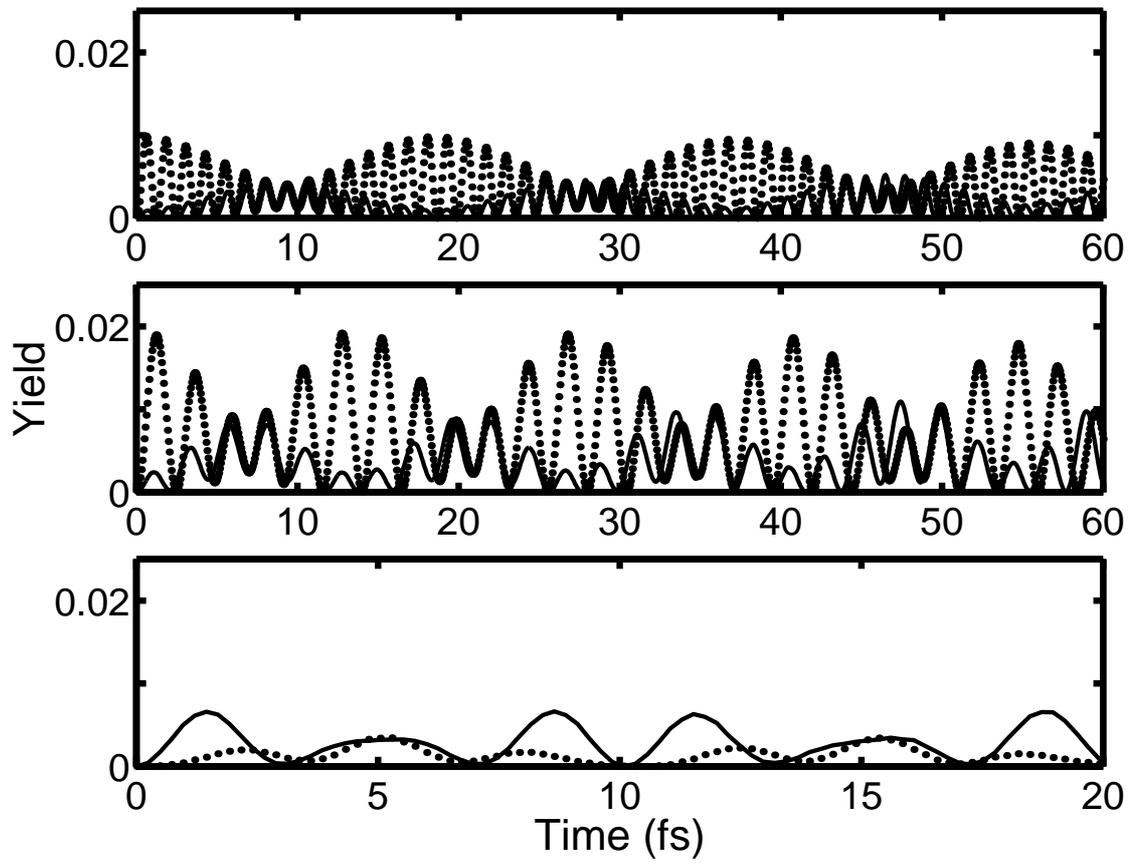
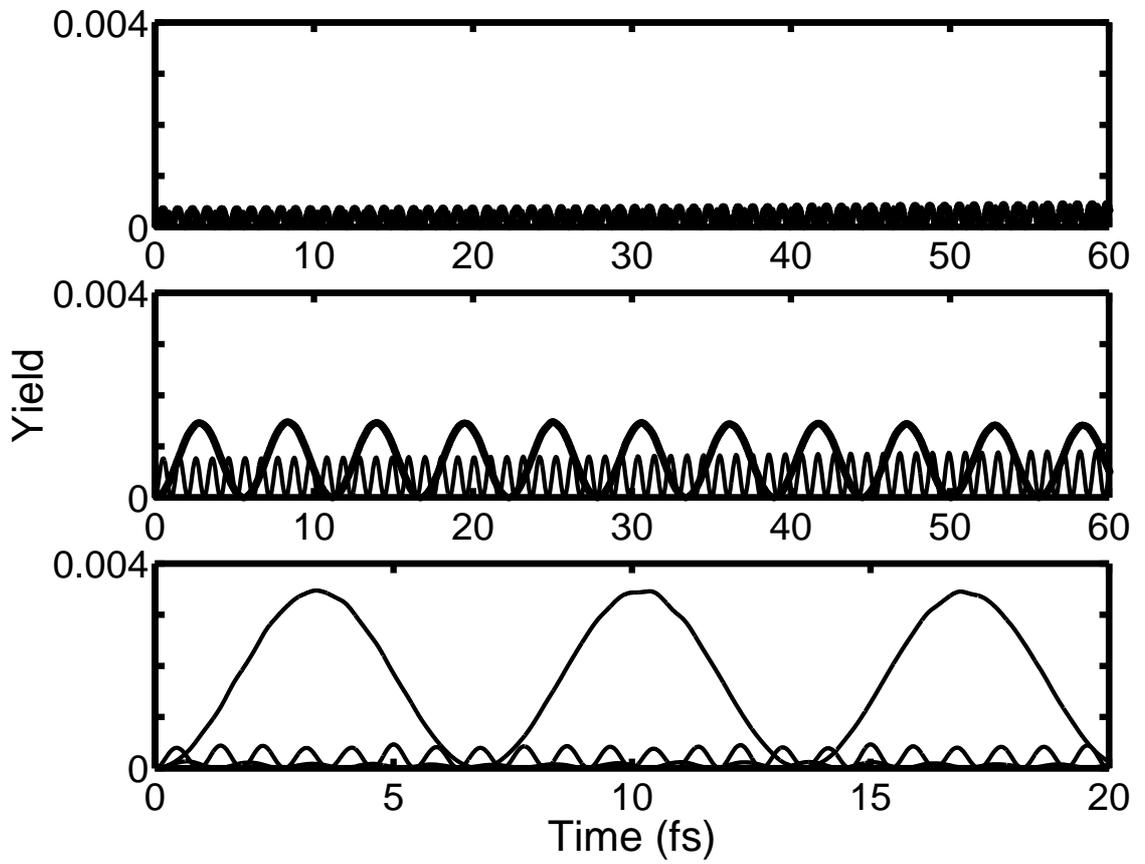
{\bf Figure 3 :} Yields in the singlet and triplet channels within the PPP Hamiltonian.
In all cases the top panel corresponds to pair of ethylenes, the middle
panel to pairs of butadienes, and the bottom panel to pairs of hexatrienes.
(a) Singlet channel, $t_{\perp}=0.1{\rm eV},~X_{\perp}=0$; (b) Triplet
channel, $t_{\perp}=0.1{\rm eV},~X_{\perp}=0$; (c) Singlet channel,
$t_{\perp}=0.1{\rm eV},~X_{\perp}=0.1 {\rm eV}$; (d) Triplet channel
$t_{\perp}=0.1{\rm eV},~X_{\perp}=0.1 {\rm eV}$.  Evolution
in case of hexatrienes is tracked for 20 fs while in other cases, the
evolution is tracked for 60fs. Significant yields in singlet channel
occurs only for final states $|(1^1A_g^+)_1(1^1B_u^-)_2>$ and
$|(1^1B_u^-)_1(1^1A_g^+)_2>$, between which the yields are identical
in (a) and (b) but different in (c) and (d). Similarly, yields in triplet
channel are to the states $|(1^1A_g^+)_1(1^3B_u^+)_2>$ and
$|(1^3B_u^+)_1(1^1A_g^+)_2>$, between which the yields are identical
in (a) and (b) but different in (c) and (d).

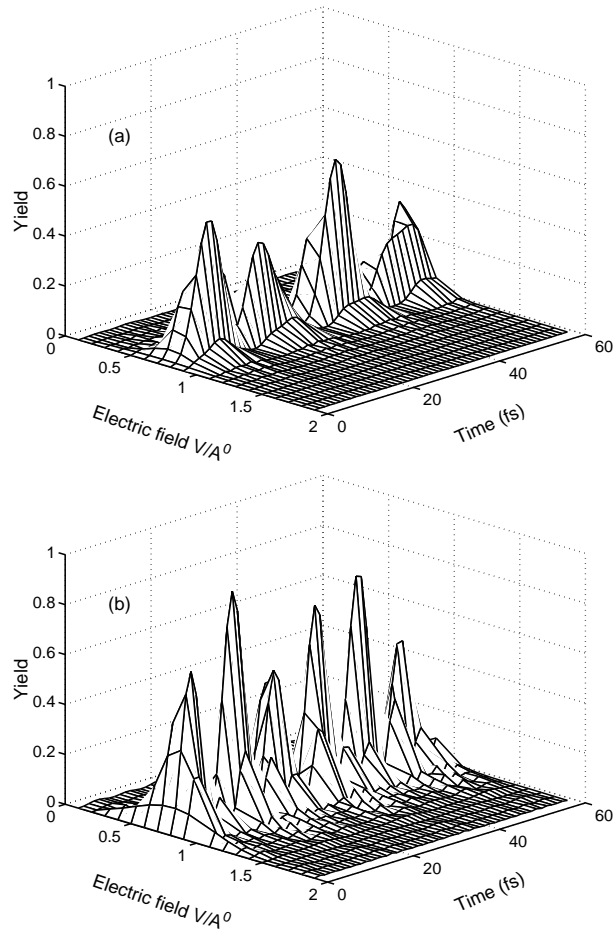
{\bf Figure 4 :} Yields in the singlet channels (a) $|(1^1A_g)_1(1^B_u)_2\rangle$, (b)
$| (1^1A_g)_2(1^B_u)_1\rangle$, as a
function of the electric field (V/\AA)
and time (fs) . Here $t_{\perp}$ = 0.1 eV and $X_{\perp}$ = 0.1 eV.

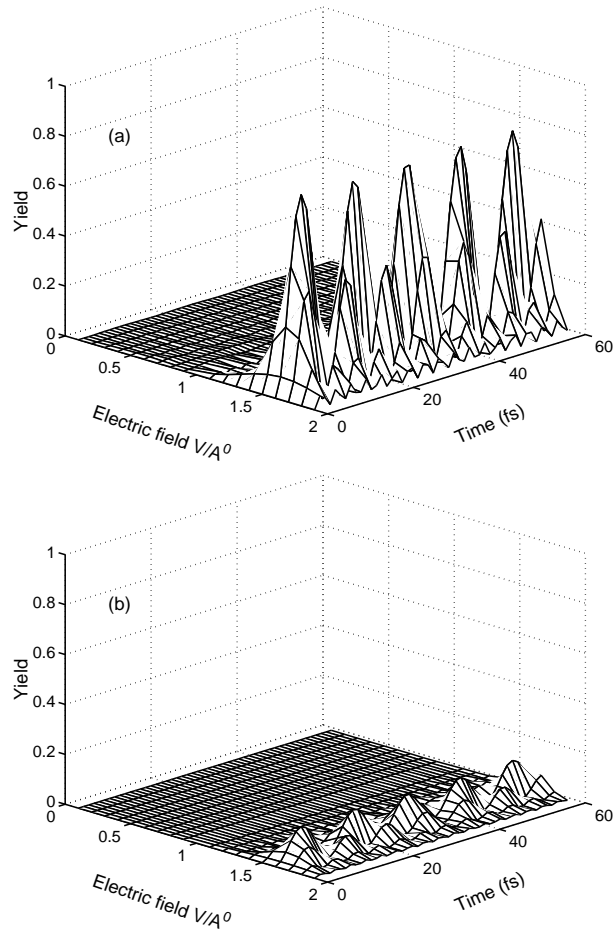
{\bf Figure 5 :} Yields in the triplet channels (a) $|(1^1A_g)_1(3^B_u)_2\rangle$, (b)
$| (1^1A_g)_2(3^B_u)_1\rangle$, as a
function of the electric field (V/\AA)
and time (fs). Parameters are same as in Fig. 4.

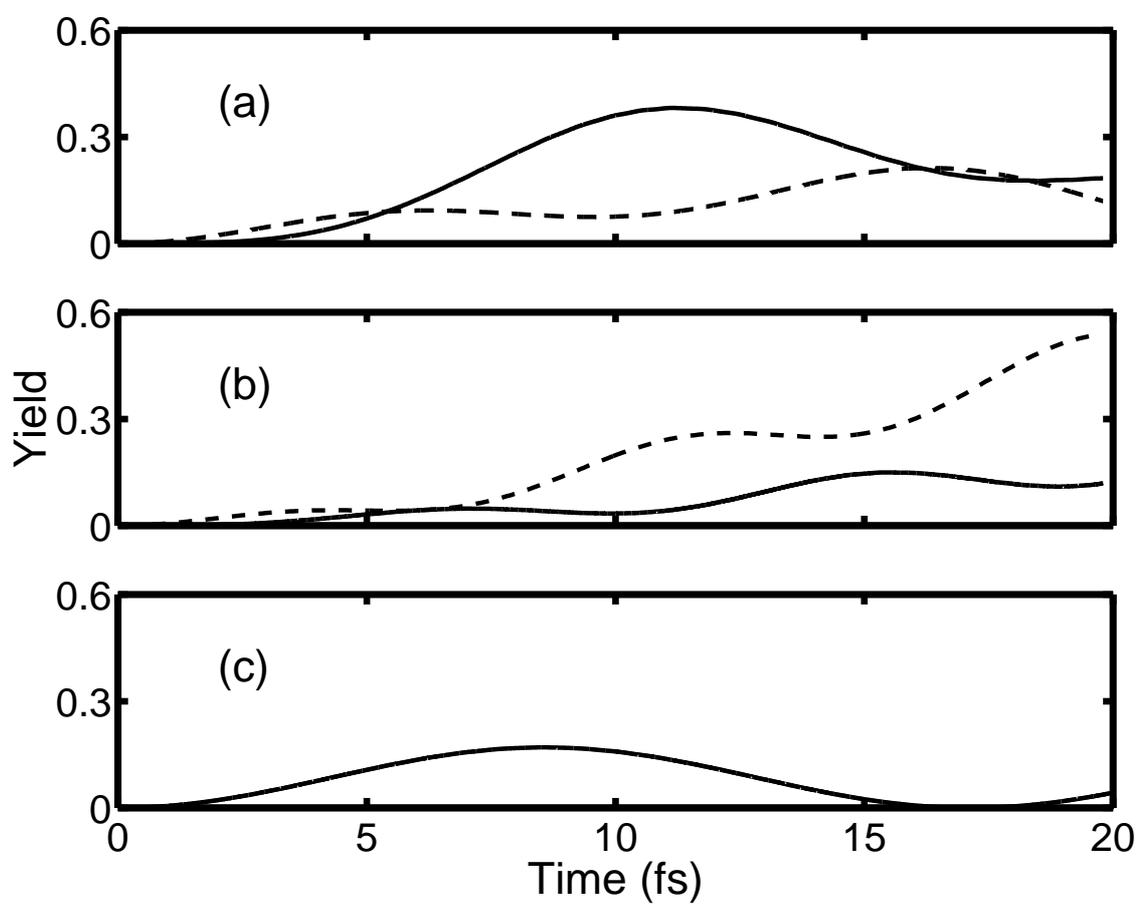
{\bf Figure 6 : } Yields in the singlet channel for pairs of hexatriene molecules,
as a function of time (fs) with $t_{\perp}=0.1$eV and $X_{\perp}=0.1$eV in
an external electric field. (a) Singlet channel at 0.3 V/\AA, (b) singlet
channel at 0.42 V/\AA and triplet channel at 1.0 V/\AA. 

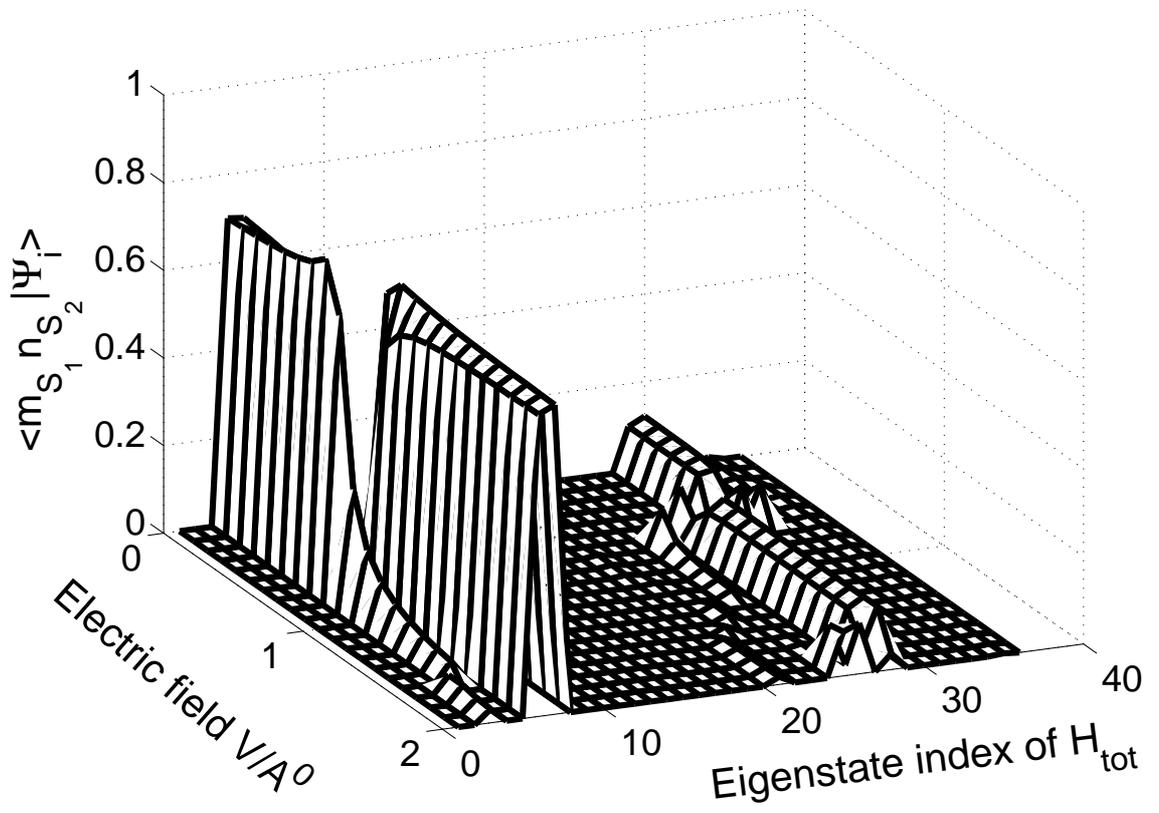
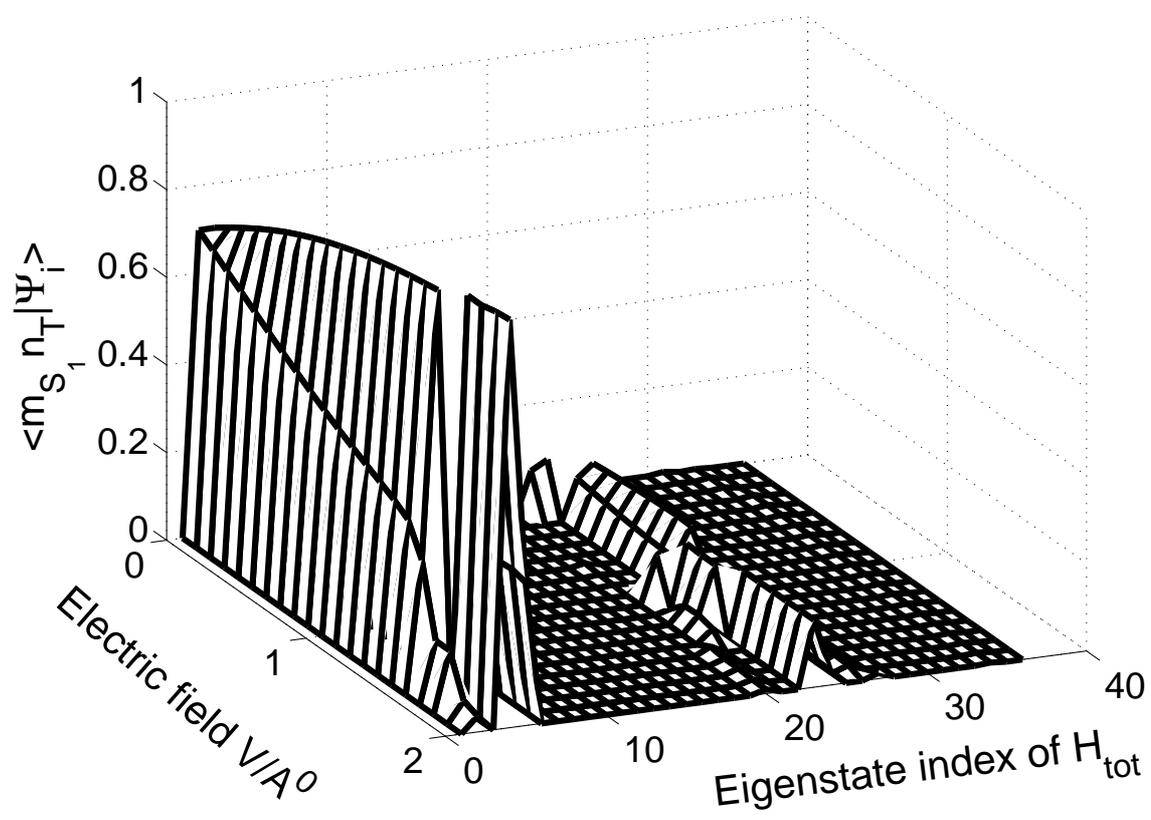
{\bf Figure 7 :} Evolution of significant $<\psi_i| m,n>$
  as a function of electric field (V/\AA), in case of the explicit
  time evolution of eigenvectors the PPP Hamiltonian for a pair of ethylenes
  in (a) singlet and (b) triplet channels.

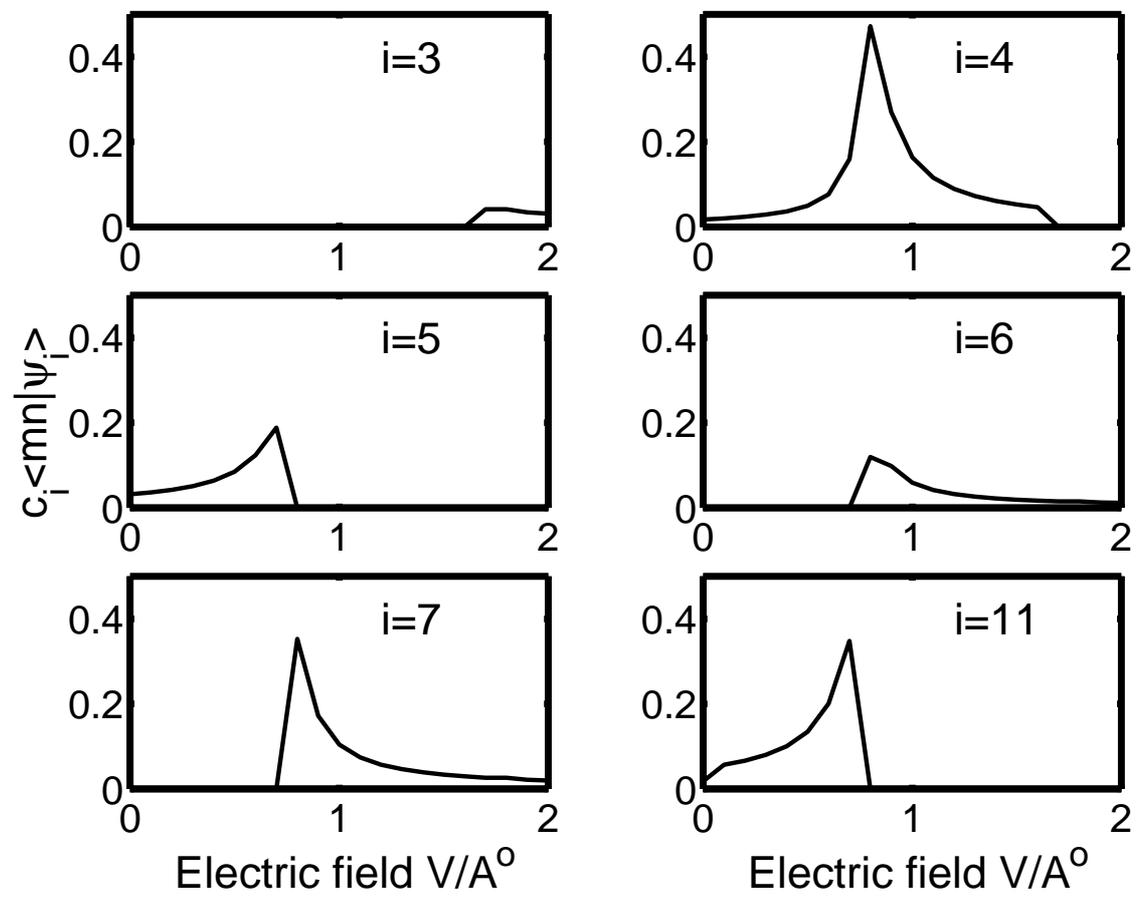
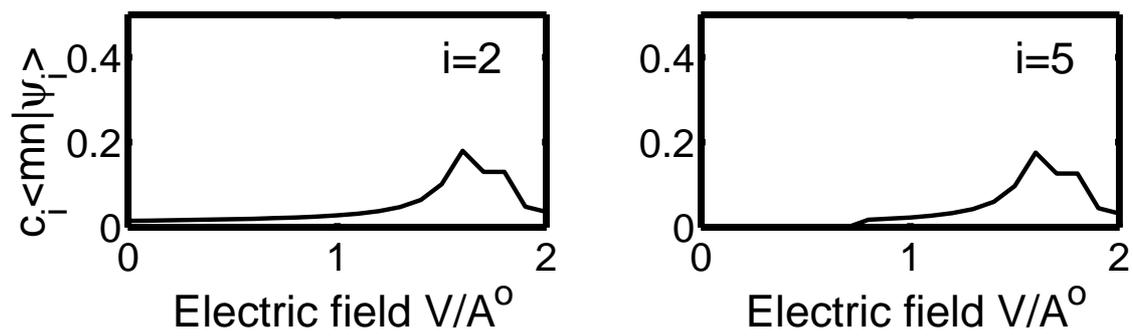
{\bf Figure 8 :} $<mn|\psi_i><\Psi(0)|\psi_i>$ plotted
  as a function of electric field (V/\AA), for significant states
  'i' for (a) the singlet-singlet channel and (b) the singlet-triplet
  channel for a pair of ethylenes. The singlet-singlet channel in
  (a) corresponds to  $|m_{S_1}>$ and $|n_{S_2}>$ and  the
  singlet-triplet channel in (b) corresponds to $|m_{S_1}>$
  and $|n_{T}>$.

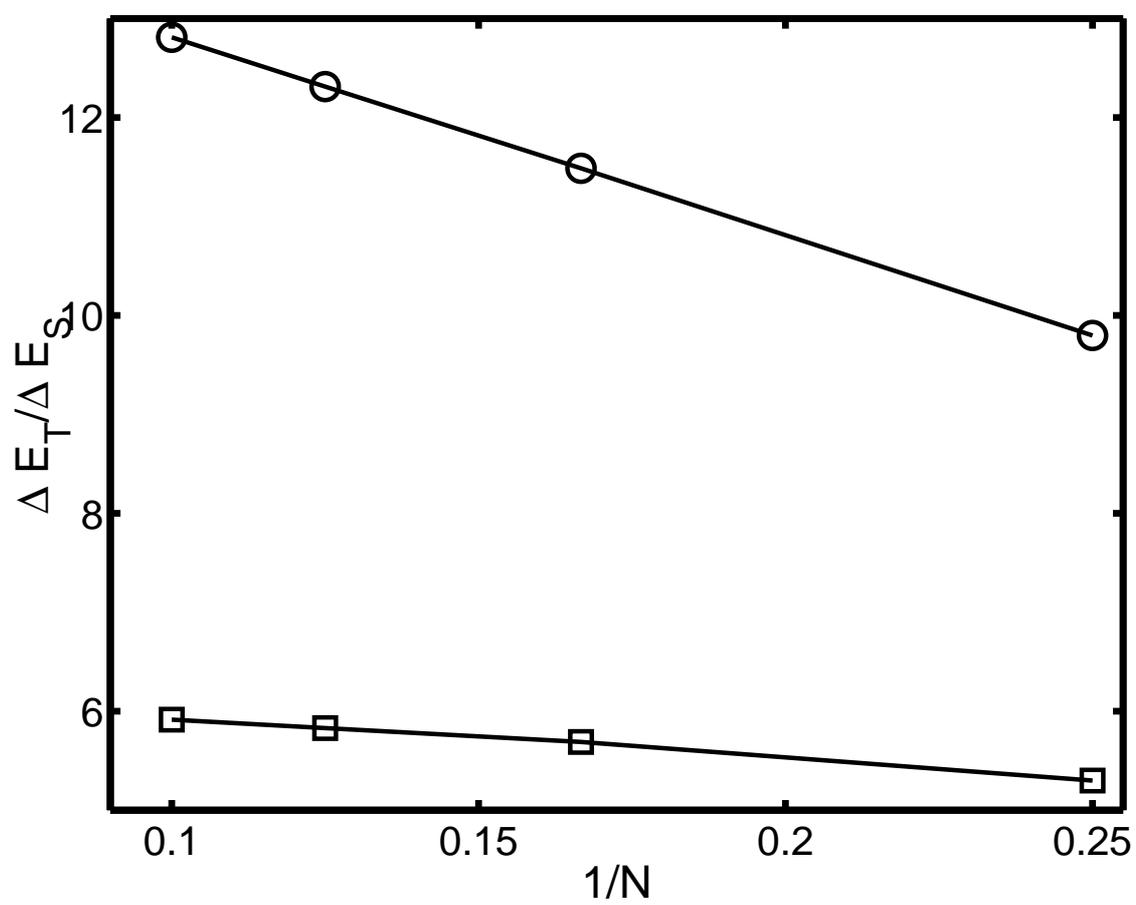
{\bf Figure 9 :} $\Delta E_T/\Delta E_S$ {\it vs} $1/N$ for the case of linear chains
with "U-V" model Hamiltonian for the case of (i) U=5eV and V=2eV (squares)
and (ii) U=6eV and V=2eV (circles).

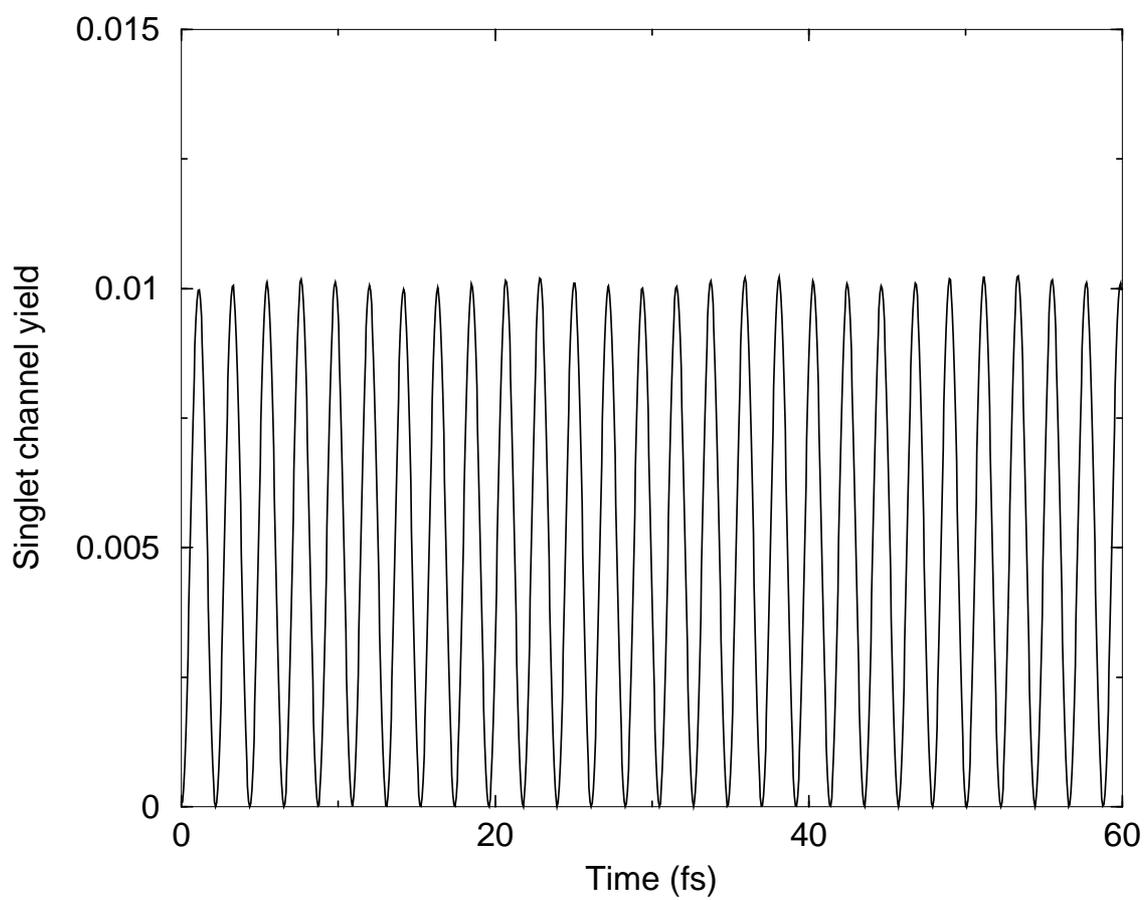
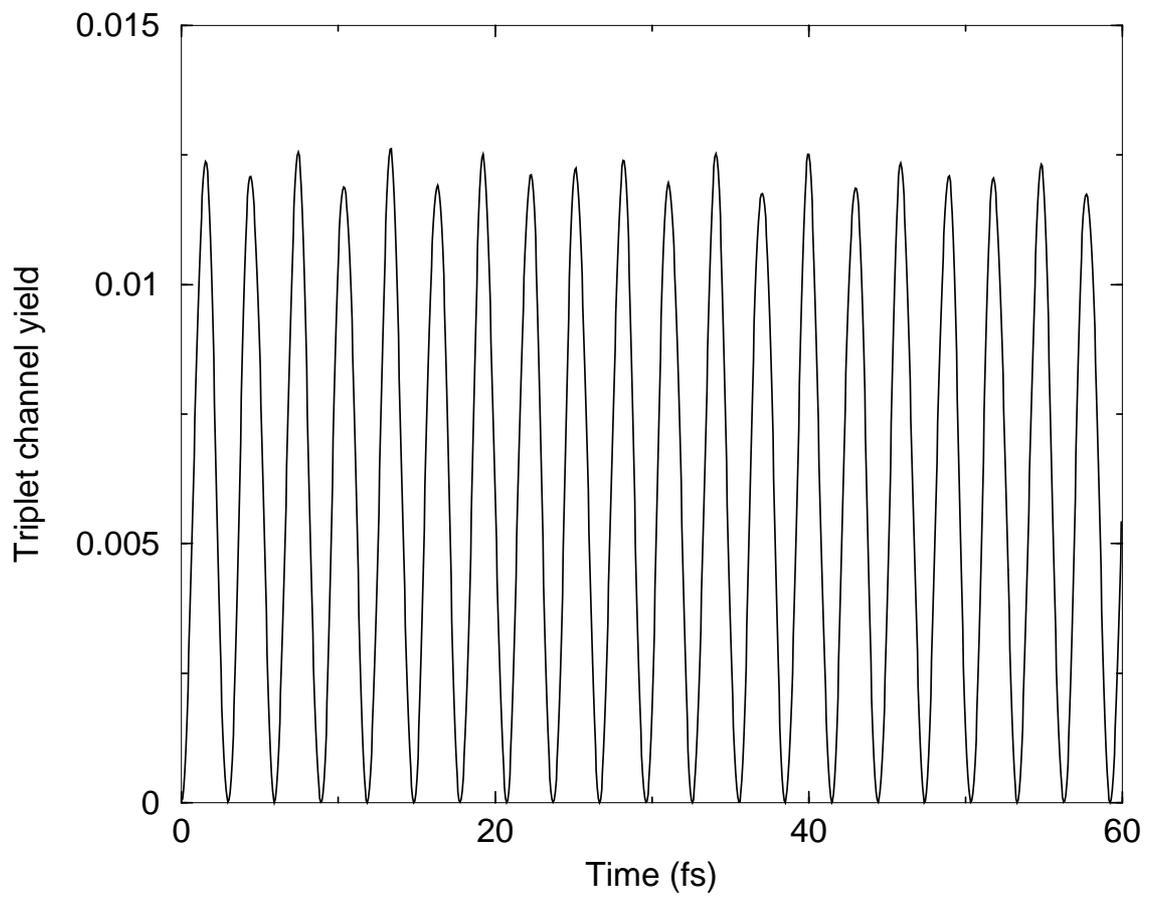
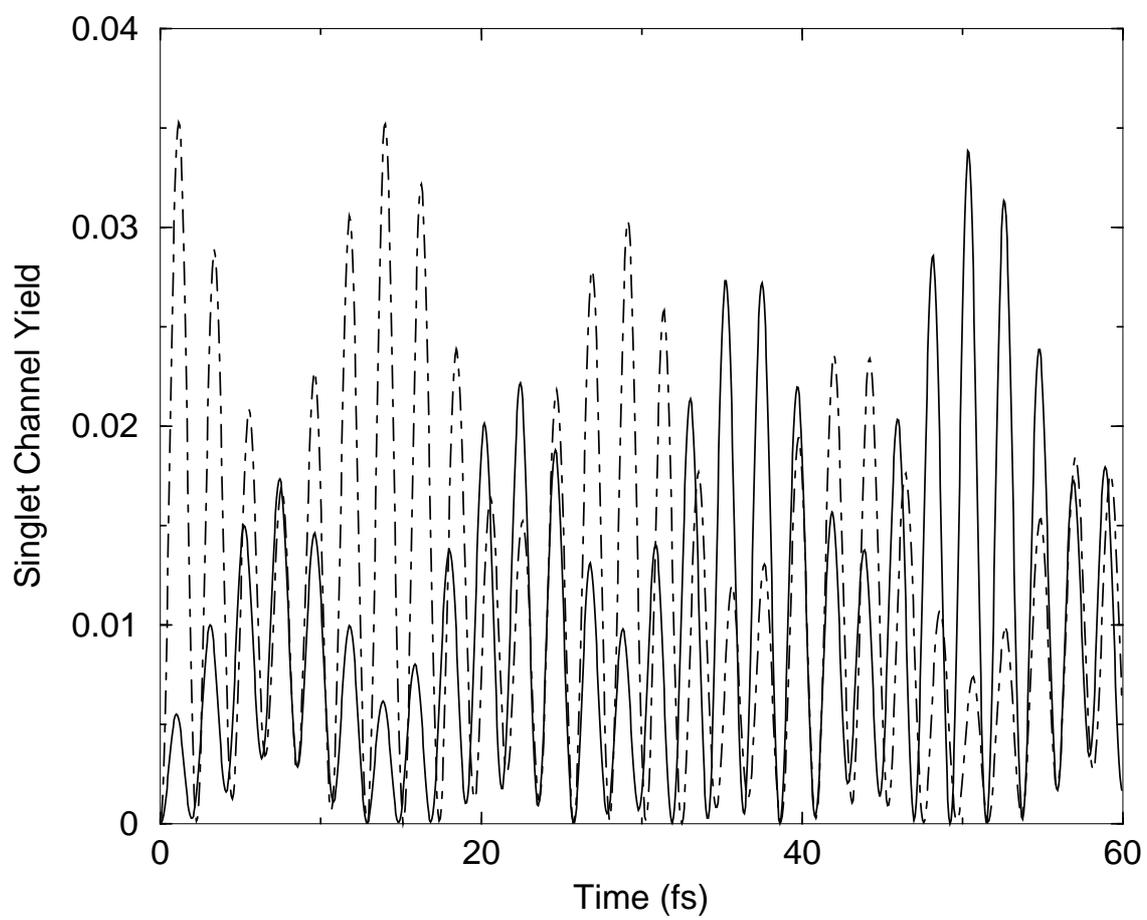
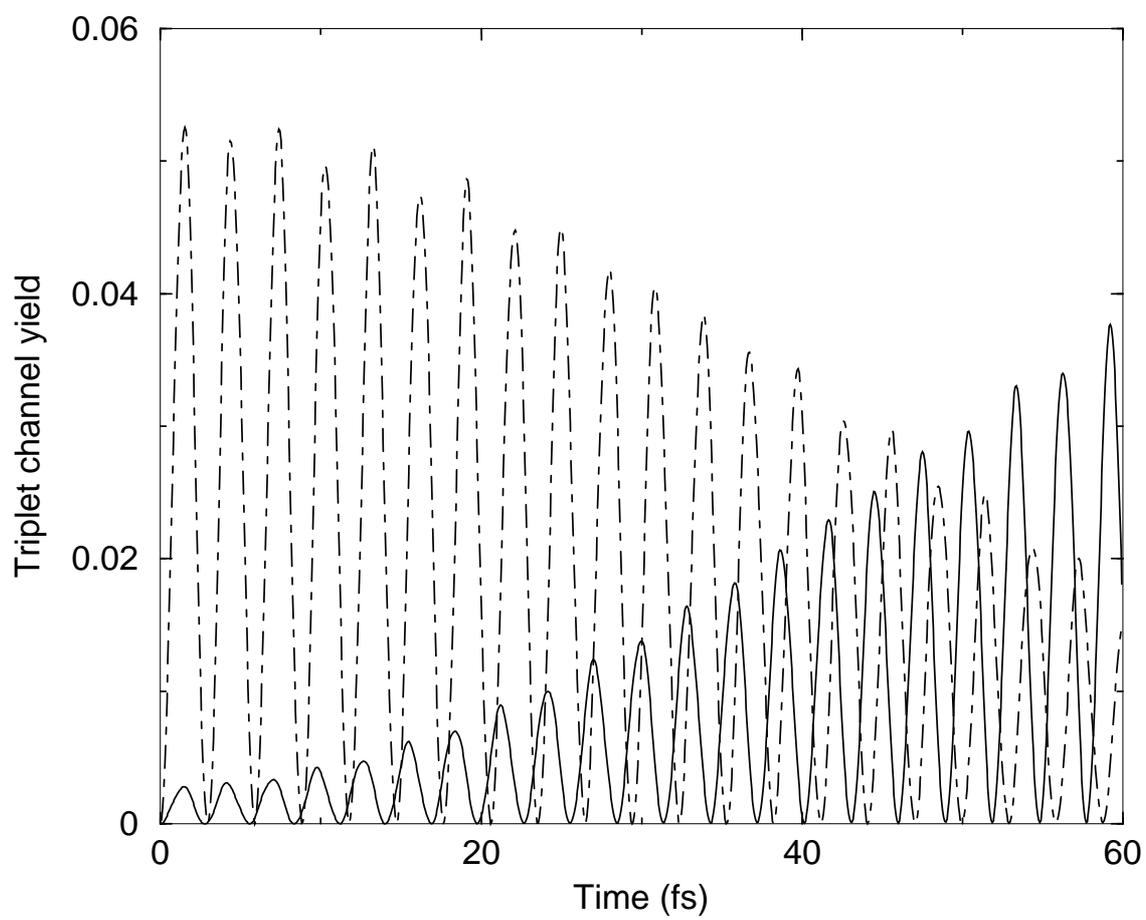
{\bf Figure 10 :} Yields in the PPP model for the (CH=N)$_2$ system. (a) singlet and (b)
triplet channels with $X_{\perp}=0$; (c) and (d) singlet and triplet
channels with $X_{\perp}=0.1$eV. The state to which the yield is
significant in (a) is $|S_0S_1>$ while in (b) it is to the state $|S_0T>$.
The yield to states $|S_1S_0>$ in singlet channel and $|TS_0>$ in
triplet channel are identical to those for $|S_0S_1>$ and $|S_0T>$
in (a) and (b) respectively. In (c) the yields to $|S_1S_0>$ and   
$|S_0S_1>$ are not the same and are shown separately. Similarly,
in (d) yields to $|TS_0>$ and $|S_0T>$ are shown separately.

{\bf Figure 11 :} The initial (a) singlet and (b) triplet states
$|P_1^+P_2^-\rangle$ for the case of two ethylenes, and the result of
operating with $H_{inter}$. The upper (lower) two sites correspond to
molecule 1 (molecule 2). The result (c) is a linear relationship between
covalent triplet VB diagrams.

\end{document}